\documentclass[12pt]{article}
\usepackage{epsfig}
\usepackage{amsfonts}
\usepackage{latexsym}
\usepackage{amsmath,amssymb}
\usepackage{verbatim}
\usepackage{setspace}
\usepackage[small]{caption}
\usepackage{graphicx}
\usepackage{color}
\usepackage{pdfsync}
\definecolor{darkgreen}{rgb}{0,0.5,0}
\definecolor{darkblue}{rgb}{0,0,0.6}
\definecolor{purple}{rgb}{0.4,0.15,0.21}
\usepackage[colorlinks=true,citecolor=darkgreen,linkcolor=purple,urlcolor=darkblue]{hyperref}

\DeclareMathOperator{\sgn}{sgn}
\newcommand{\f}{\frac}
\newcommand{\reals}{\mathbb{R}}

\usepackage[textheight=9in, textwidth=6.5in, letterpaper]{geometry}

\numberwithin{equation}{section}

 \def\p{\partial}

\def\tilde{\widetilde}
\newcommand{\bea}{\begin{eqnarray}}
\newcommand{\eea}{\end{eqnarray}}
\newcommand{\be}{\begin{equation}}
\newcommand{\ee}{\end{equation}}
\newcommand{\ba}{\begin{align}}
\newcommand{\ea}{\end{align}}

\renewcommand{\epsilon}{\varepsilon}

  \makeatletter
  \let\over=\@@over \let\overwithdelims=\@@overwithdelims
  \let\atop=\@@atop \let\atopwithdelims=\@@atopwithdelims
  \let\above=\@@above \let\abovewithdelims=\@@abovewithdelims

\renewcommand\section{\@startsection {section}{1}{\z@}%
                                   {-3.5ex \@plus -1ex \@minus -.2ex}
                                   {2.3ex \@plus.2ex}%
                                   {\normalfont\large\bfseries}}

\renewcommand\subsection{\@startsection{subsection}{2}{\z@}%
                                     {-3.25ex\@plus -1ex \@minus -.2ex}%
                                     {1.5ex \@plus .2ex}%
                                     {\normalfont\bfseries}}

\newcommand{\Tr}{\mbox{Tr}}

\def\ep{\epsilon}

\def\Or[#1]{{\text{O}}\left({#1}\right)}
\def\dotl[#1,#2]{\left\langle #1, #2 \right\rangle}
\def\dotlb[#1,#2]{[ #1, #2 ]}
\def\dotp[#1,#2]{(#1) \cdot (#2)}
\def\aff[#1,#2]{\hat{#1}(#2)}
\def\n4sym{{\cal N}=4 SYM}
\def\>{\rangle}
\def\<{\langle}
\def\weight[#1,#2,#3]{\{(#1),#2,#3\}}
\def\ads[#1]{$\text{AdS}_{#1}$}

\def\p{\partial}

\begin{document}
\unitlength = 1mm

\begin{flushright}
SU-ITP-13/18
\end{flushright}
\ \\
\begin{center}

{ \LARGE {\textsc{Warped Entanglement Entropy}}}

\vspace{0.8cm}

Dionysios Anninos$^1$, Joshua Samani$^{2}$ and Edgar Shaghoulian$^1$

\vspace{1cm}

\vspace{0.5cm}

$^1$ {\it Stanford Institute for Theoretical Physics, Stanford University}\\

$^2$ {\it Department of Physics and Astronomy, University of California, Los Angeles}\\

\vspace{1.0cm}

\end{center}

\begin{abstract}
  We study the applicability of the covariant holographic entanglement entropy proposal to asymptotically warped AdS$_3$ spacetimes with an $SL(2,\mathbb{R})\times U(1)$ isometry. We begin by applying the proposal to locally AdS$_3$ backgrounds which are written as an $\mathbb{R}^1$ fibration over AdS$_2$. We then perturb away from this geometry by considering a warping parameter $a=1+\delta$ to get an asymptotically warped AdS$_3$ spacetime and compute the dual entanglement entropy perturbatively in $\delta$. We find that for large separation in the fiber coordinate, the entanglement entropy can be computed to all orders in $\delta$ and takes the universal form appropriate for two-dimensional CFTs. The warping-dependent central charge thus identified exactly agrees with previous calculations in the literature. Performing the same perturbative calculations for the warped BTZ black hole again gives universal two-dimensional CFT answers, with the left-moving and right-moving temperatures appearing appropriately in the result.
\end{abstract}

\pagebreak
\setcounter{page}{1}
\pagestyle{plain}

\setcounter{tocdepth}{1}

\tableofcontents

\section{Introduction}

Understanding how the holographic principle works beyond the example of anti-de Sitter space is a crucial and beautiful challenge which will elucidate the dynamics of quantum gravity in general backgrounds. A natural example is the geometry describing our universe, which is cosmological in nature, and more closely resembles an FRW/de Sitter type universe. As another example, the geometry describing regions near the horizons of certain astrophysical black holes is not quite anti-de Sitter space but more closely resembles a slight deformation thereof known as the NHEK/warped AdS$_3$ geometry. There have been several proposals for holographic descriptions of these and other non-AdS spacetimes \cite{ds,nhek,schro}, and the story is still unfolding.

In this paper we will focus on aspects of the warped AdS$_3$ geometry and its putative holographic description. As we will describe more concretely below, warped $\mathrm{AdS}_3$ is a deformation of AdS$_3$ that destroys the boundary asymptotics. The deformation preserves only an $SL(2,\mathbb{R}) \times U(1)$ subgroup of the original  $SL(2,\mathbb{R}) \times SL(2,\mathbb{R})$ isometry group of AdS$_3$. From the point of view of the two-dimensional CFT dual to AdS$_3$, the warping of AdS$_3$ corresponds to an irrelevant chiral deformation (which does not die away in the ultraviolet). Geometrically this manifests itself in the destruction of an asymptotically AdS$_3$ boundary. Holographic considerations of this geometry began with \cite{nhek}. Based on the thermodynamic properties of asymptotically warped AdS$_3$ black holes \cite{Nutku:1993eb,gurses,Moussa:2003fc,Bouchareb:2007yx}, whose entropy could be written in a suggestive, Cardy-like fashion, it was proposed that it was dual to a two-dimensional conformal(-esque) field theory. Later work embedded and studied warped AdS$_3$ within string theory \cite{Anninos:2008qb,Orlando:2010ay,Song:2011sr,Detournay:2010rh,ElShowk:2011cm,Azeyanagi:2012zd, Karndumri:2013dca} 
and studied properties of two-dimensional field theories, dubbed warped CFTs, whose symmetry structure matches that of warped AdS$_3$ \cite{Hofman:2011zj,Detournay:2012pc}. Other work studied the wave equation, correlation functions and quasinormal modes of fields in warped AdS$_3$ \cite{Chen:2009rf,Chen:2009hg,Chen:2009cg,Anninos:2009jt,Anninos:2010gh}. Much of the work on warped AdS$_3$ has so far focused on thermodynamic properties of the theory and its asymptotic symmetry structure \cite{Compere:2008cv,Compere:2009zj}. In this paper we would like to focus instead on entangling properties of asymptotically warped AdS$_3$ geometries. We do so by exploiting the simple holographic manifestation of the entropy of entanglement of some state in a CFT as an extremal surface in the bulk geometry dual to such a state, as described by \cite{Hubeny:2007xt}, generalizing  \cite{Ryu:2006bv,Ryu:2006ef}. Though entanglement entropy is a simple property of the quantum state, it has sufficient information to independently verify features derived from the thermodynamics, such as central charges and left- and right-moving temperatures.  It can also provide additional insight into the nature of the dual as we will shortly discuss. We now move on to briefly review the warped AdS$_3$ geometry and the holographic entanglement entropy proposal before summarizing our results and giving an outline of the paper.

\subsection{Warped $\mathrm{AdS}_3$}

\begin{figure}[tb]
\centering
\includegraphics[scale=0.3]{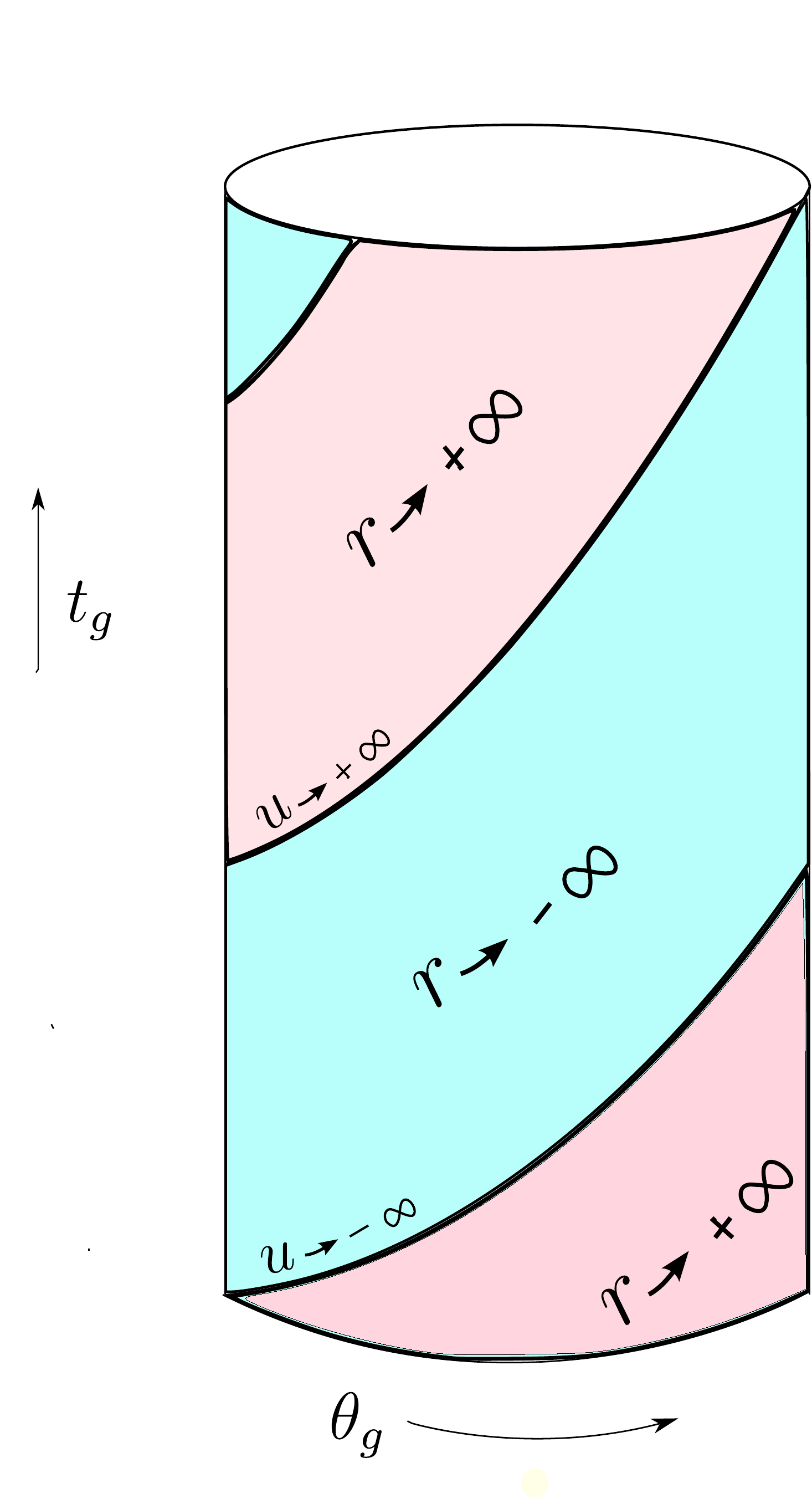}
\caption{This is the global AdS$_3$ cylinder parameterized by the coordinates \eqref{globalfibered}. The coordinates $t_g$ and $\theta_g$ represent the usual global coordinates. We will primarily consider sticking to a region of the boundary with $r=\infty$ for simplicity. This figure is taken from \cite{Anninos:2009zi}.}\label{ads3cylinder}
\label{penrose}
\end{figure}

Consider AdS$_3$ expressed as a real-line or circle fibration over a Lorentzian AdS$_2$ base space. These geometries can be deformed with a nontrivial warp factor into the warped AdS$_3$ spacetimes we will consider later. The (spacelike) warped $\mathrm{AdS}_3$ metric in global coordinates with warp factor $a\in[0,2)$ is given by\footnote{In the literature, usually in the context of topologically massive gravity, one often sees an alternative convention in which the metric is characterized by parameters $\tilde{\ell}$ and $\nu$ related to our parameters by $\tilde{\ell}^2 = \ell^2(\nu^2+3)/4$ and $a^2 = 4\nu^2/(\nu^2+3)$.}
\begin{align}
    ds^2 = \frac{\ell^2}{4}\left(-(1+r^2)\,d\tau^2 + \frac{dr^2}{1+r^2}+a^2(du + r\,d\tau)^2\right)\label{globalfibered}.
\end{align}
The coordinates range over the whole real line, $\{r,\tau,u\} \in \mathbb{R}^3$, although later we will consider compactifying $u$ to recover a near-horizon extremal BTZ geometry. To obtain $\mathrm{AdS}_3$, one sets $a=1$. The conformal boundary in the case of $a=1$ is the usual cylinder parsed by null coordinates and looks like a barber-shop pole; see Figure \ref{ads3cylinder}. The case $a\neq1$ corresponds to spacelike warped AdS$_3$, which is the case we shall focus on in this paper. We will also comment on the timelike warped AdS$_3$ case, whose base space is Euclidean AdS$_2$, in Section \ref{vacuum}. For $a \neq 1$ there is no conformal boundary \cite{Bengtsson:2005zj}, although a generalized notion of ``anisotropic conformal infinity" can be defined \cite{Horava:2009vy}.  We will also consider the geometries in Poincar\'e-like coordinates with metric
\begin{align}
    ds^2 = \f{1}{4}\left(-\ell^2\f{d\psi^2}{x^2} + \ell^2\f{dx^2}{x^2}+a^2\left(d\phi + \ell\f{d\psi}{x}\right)^2\right)\label{poinfibered}.
\end{align}
and coordinate ranges $\{\psi,x,\phi\}\in \mathbb{R}^3$. 

These spacetimes posses ${SL}(2,\reals) \times U(1)$ isometry for $a\neq 1$ and appear in a Penrose-like near-horizon limit of extremal black holes. In the context of a trivial warp factor $a=1$, these geometries are locally AdS$_3$, and we expect the HRT proposal to apply. We will see that our results match field theory expectations, where the field theory is placed at zero left-moving temperature and finite right-moving temperature. This state of the field theory has not yet been considered in the holographic entanglement entropy literature, though it is closely related to the extremal limit of the rotating BTZ black hole, considered in \cite{Hubeny:2007xt}.

For the case of nontrivial warp factor, the purported holographic duals of the spacetime are referred to as warped CFTs and possess ${SL}(2,\reals)\times  U(1)$ symmetry. This symmetry is automatically enhanced to two infinite-dimensional local symmetries \cite{Hofman:2011zj}: the left-moving ${SL}(2,\reals)$ is enhanced to a left-moving Virasoro, while the right-moving $U(1)$ is enhanced to a left-moving $U(1)$ Kac-Moody current algebra (indeed, the term WCFT is used for the case that the $U(1)$ is \emph{not} enhanced to a full Virasoro, which is also possible). Not much is known about these theories (a nontrivial example has only recently been suggested in \cite{Compere:2013bya}), but the symmetries can still be used to constrain properties that such a theory could have. This approach has been used successfully in reproducing a Cardy-like formula for the asymptotic growth of states in \cite{Detournay:2012pc}. The bulk geometries are often considered in the context of topologically massive gravity, but for simplicity we shall restrict ourselves to the case where they are solutions of three-dimensional Einstein gravity with matter fields, as studied in \cite{Duff:1998cr,Anninos:2008qb, Detournay:2010rh}. In string theory, for example, the warped geometries can be constructed by a hyperbolic, marginal deformation of the ${SL}(2,\mathbb{R})$ WZW model \cite{Israel:2004vv}.

\subsection{Holographic entanglement}

The use of entanglement entropy to study quantum field theories continues to surge due to its relevance to quantum gravity and condensed matter physics and its analytic tractability. Holographically, this has been studied with the Ryu-Takayanagi (RT) proposal \cite{Ryu:2006bv, Ryu:2006ef} for computing the entanglement entropy via geometric methods in the bulk. The proposal now has  support for multiple intervals in asymptotically AdS$_3$ bulk spacetimes \cite{Headrick:2010zt, Hartman:2013mia, Faulkner:2013yia} and spherical entangling surfaces in any dimension \cite{Casini:2011kv}. Strong arguments for the general case are provided in \cite{Lewkowycz:2013nqa} and essentially prove the conjecture. Quantum corrections have been analytically computed in \cite{Barrella:2013wja}, with a general prescription appearing in \cite{Faulkner:2013ana}. Prescriptions for gravitational theories with higher curvature corrections are given in \cite{Hung:2011xb, Bhattacharyya:2013jma, Fursaev:2013fta, Bhattacharyya:2013gra, Sun:2008uf} and, for higher spin theories, in \cite{Ammon:2013hba, deBoer:2013vca}. The covariant Hubeny-Rangamani-Takayanagi (HRT) proposal \cite{Hubeny:2007xt} has far less support, though it has passed nontrivial consistency checks \cite{Callan:2012ip, Wall:2012uf}. It is natural to wonder how generally the proposal can apply. In this paper, we would like to take a few steps toward understanding the issues of holographic entanglement entropy in warped AdS$_3$ spacetime and two-dimensional warped conformal field theory (WCFT$_2$). The spacetimes we will study are non-static and will therefore require the covariant proposal. Although these spacetimes are often studied as solutions of topologically massive gravity, here we will consider the case where they are supported by Einstein gravity plus matter, allowing us to use the usual HRT proposal.

The goal of the HRT proposal in \cite{Hubeny:2007xt} is to obtain a holographic prescription for computing the entanglement entropy for time-varying states in QFTs with bulk duals which are non-static, asymptotically AdS spacetimes.  To describe the proposal, we consider a $(d+1)$-dimensional asymptotically AdS spacetime $M$ with $d$-dimensional boundary $\partial M$, and we consider a field theory defined on this boundary.  We choose a foliation of $\partial M$ by spacelike hypersurfaces (time slices) $\widehat M_t$ .  For each time $t\in\reals$, we write the slice $\widehat M_t$ as a union of disjoint sets $A_t$ and $B_t$, and we can compute the entanglement entropy  $S_{AB}(t)$ between the degrees of freedom in the two regions for a given state (density matrix) of the full system living on $\widehat M_t$.

The HRT proposal is as follows:  for each time $t$, determine the co-dimension 2 extremal surfaces $W_t$ satisfying $\partial W_t = \partial A_t$.  If there is more than one extremal surface satisfying these boundary data, then choose the extremal surface $W_{t, \mathrm{min}}$ with smallest area. Then we have
\begin{align}\label{hrt}
    S_{AB}(t) = \frac{\mathrm{Area}(W_{t,\mathrm{min}})}{4 G_N^{(d+1)}}.
\end{align}
The question of which homology class to consider is interesting in the context of the covariant proposal \cite{Hubeny:2013gta}, but we will not need to consider it here. This is the correct expression for Einstein gravity coupled to matter, which are the theories we will consider here, although subleading corrections in $G_N$ (bulk quantum corrections) will depend on the bulk matter supporting the geometry.

It is worth noting a rather remarkable feature of the above proposal (\ref{hrt}). In the context of Einstein theories of gravity the entanglement entropy manifests itself in a purely geometric form at leading order in $G_N$, as the area of an extremal surface. This universal feature is particularly surprising, given that entanglement entropy is a property of the particular quantum state under consideration, which is generally a functional of all the bulk matter fields and not just the metric. It is reminiscent of the universality of the Bekenstein-Hawking entropy of a black hole, which also manifests itself as a geometric area in Einstein theories of gravity, regardless of the matter content that constitutes the black hole.

In our use of this formula, we will keep the slice chosen on the boundary arbitrary but spacelike. Since we consider exclusively (2+1)-dimensional bulk geometries, this means that our entanglement entropy answers will be phrased in terms of two distinct coordinate separations, which can then be chosen to give a particular spacelike slice. We present the answers in this way because it makes the split into left-moving and right-moving sectors transparent; see \eqref{btzsectors} for one such example. It is important to note that the HRT prescription (and indeed the original Ryu-Takayanagi prescription) is computing the entanglement entropy between regions $A_t$ and $B_t$ defined by the unique geodesic along the \emph{boundary} which connects the two points which define their separation. In other words, once one picks two points on the boundary to connect by a bulk geodesic, there remains an ambiguity in choosing the spacelike curves on the boundary which connect the two points and define the regions $A_t$ and $B_t$. The holographic entanglement entropy prescription naturally picks the unique geodesic along the boundary which connects the two points as defining the spatial regions  $A_t$ and $B_t$.

To elaborate further, imagine applying the covariant proposal to the Poincar\'e patch of AdS$_3$ by picking points on the boundary that are spacelike separated but arbitrary. The length of the regulated bulk geodesic connecting these two points, divided by $4G_N$, is given in terms of CFT quantities as
\be
S_{EE}=\f{c}{3}\,\log\f{\sqrt{L_x^2-L_t^2}}{\epsilon}\,.\label{boost}
\ee
To match with the universal 2D CFT answer, we conclude that the region being picked out on the boundary theory is the geodesic along the boundary which connects the two points, since this curve has length $\sqrt{L_x^2-L_t^2}$. The fact that in this example the spatial length at fixed time gets replaced with the invariant Minkowskian length is a result of Lorentz invariance. This will not be the case once we introduce a dimensionful scale, e.g. the radius of the cylinder for global AdS$_3$ or the temperature of a black hole.

\subsection{Summary and outline}

Although the validity of applying the HRT proposal to spacetimes with different asymptotics is an interesting open question,\footnote{We stress that calculations like the ones in \cite{Ogawa:2011bz}, which consider a decoupled IR geometry, are still understood as occurring in an asymptotically AdS spacetime, as discussed in \cite{Shaghoulian:2011aa}.} in this paper we shall pursue a more modest goal. We will set up what is effectively a perturbation theory about the AdS$_3$ point by considering warping $a=1+\delta$ and cutting off the WAdS$_3$ spacetime deep in the interior, where it is AdS$_3$-like. This can be understood as AdS/CFT in the presence of an infinitesimal, irrelevant deformation, a context in which holographic renormalization can be understood perturbatively in the deformation \cite{vanRees:2011fr, vanRees:2011ir}.\footnote{For a specific implementation in Lifshitz backgrounds with $z=1+\epsilon$, see \cite{Korovin:2013bua}.} Thus, attacking the problem in this way puts our analysis on firmer footing. We will see that such an approach gives sensible results, and in the regime of large separation in the fiber coordinate, the series can be summed to all orders in $\delta$. The result is precisely that of two-dimensional CFT, with $c_L=c_R=3\ell a/2G_N$.  This exactly matches an independent proposal for the central charge, deduced by demanding consistency with the Cardy formula for two-dimensional CFTs \cite{Anninos:2008qb}. We will also consider the warped BTZ black hole and again find universal CFT results which allow us to read off the left-moving and right-moving temperatures. Our central charge and temperatures altogether satisfy the Cardy formula and reproduce the entropy of the warped BTZ black hole.

In Section \ref{ads3sec}, we will apply the HRT proposal to locally AdS$_3$ spacetimes written as a fibration over a Lorentzian AdS$_2$ base space. In Section \ref{wads3}, we will deform these geometries into warped AdS$_3$ and set up the problem of applying the HRT proposal to these spacetimes. In Section \ref{pertsec}, we will complete the problem by performing a perturbative application of the HRT proposal to warped AdS$_3$ geometries, where we will be perturbing around the locally AdS$_3$ geometries considered in Section \ref{ads3sec}. Finally, we will summarize and look toward future work in Section \ref{summsec}. 

\section{$\mathrm{AdS}_3$ in fibered coordinates}\label{ads3sec}

We begin our story by considering AdS$_3$ in fibered Poincar\'e coordinates and fibered global coordinates. We will see that these coordinate systems are dual to states at zero left-moving and finite right-moving temperature, a feature reflected in the answer for the entanglement entropy. The geometry obtained by compactifying the fiber coordinate appears in a near-horizon limit of the extremal BTZ black hole. If the fiber coordinate remains uncompactified, the geometry is instead the near-horizon limit of a boosted extremal black string.

\subsection{Poincar\'e fibered $\mathrm{AdS}_3$}

The metric (\ref{poinfibered}) with $a=1$ reduces to
\begin{align}
  ds^2=\f{1}{4}\left(-\ell^2\frac{d\psi^2}{x^2}+\ell^2\f{dx^2}{x^2}+\left(d\phi+\ell\f{d\psi}{x}\right)^2\right).\label{fads3}
\end{align}
We choose this parameterization since all coordinates and $\ell$ can be assigned dimensions of length. Near the conformal boundary, the coordinates $\phi$ and $\psi$ become null. We would like to determine the affinely parametrized geodesics $x^\mu(\lambda)=(x(\lambda),\phi(\lambda),\psi(\lambda))$. To do so, we notice that this geometry has Killing vectors $\partial_\phi$ and $\partial_\psi$, corresponding to translations in $\phi$ and $\psi$, and these Killing vectors yield conserved quantities $c_\phi = \dot x\cdot \partial_\phi$ and $c_\psi = \dot x\cdot \partial_\psi$. We will solve for the geodesics by using these conserved quantities and the affine constraint $c_v = \dot x^\mu \dot x_\mu$. This gives equations of motion
\begin{align}
    c_\phi &= \frac{x \dot \phi+\ell\dot \psi}{4x}\;, \label{pcphi}\\
    c_\psi &= \frac{\ell\dot \phi}{4x}\;, \label{pcpsi}\\
    c_v &= \frac{\ell^2 \dot x^2+x\dot \phi(x\dot \phi+2\ell \dot \psi)}{4x^2}\;. \label{pcE}
\end{align}
The solutions to these equations are given in Appendix \ref{poinadssols}.

We want to compute the length of a geodesic beginning and ending near the conformal boundary at $x_\epsilon\sim\epsilon^2/\ell$, where we have used the UV-IR relation to map our bulk IR cutoff $x_\epsilon$ to a dual UV field theory cutoff $\epsilon$ \cite{Susskind:1998dq, Ryu:2006ef}. This follows from the quadratic relationship between $x$ and the usual Poincar\'e coordinate $z$ near the conformal boundary, $x\sim z^2$. Since we chose our geodesics to be affinely parametrized, we can use the solution $x(\lambda)$ to solve for the cutoffs $\pm \lambda_\infty$ in the affine parameter defined by $x(\pm\lambda_\infty)=x_\epsilon$. The regulated length is then given by
\be
  \textrm{Length}=\sqrt{c_v}\int_{-\lambda_\infty}^{\lambda_\infty}d\lambda=2\sqrt{c_v}\;\lambda_\infty,
\ee
which at the end will be $c_v$-independent, as required by parameterization-invariance of the length. Writing down the leading divergence of $\lambda_\infty$ in terms of the conserved quantities $c_\phi$, $c_\psi$ and $c_v$ gives
\be
  \lambda_\infty\approx \f{\ell}{4\sqrt{c_v}}\log\left(\f{c_v\,\ell\sqrt{4c_\phi^2-c_v}}{4(c_vc\psi-4c\phi^2c\psi)}\;\f{\ell}{\epsilon}\right),
\ee
we now attempt to trade the conserved quantities $c_\phi$ and $c_\psi$ for spatial separations on the asymptotic boundary. This entails solving the equations
\begin{align}
  L_\phi\equiv\phi(\lambda_\infty)-\phi(-\lambda_\infty)\;,\qquad
  L_\psi\equiv\psi(\lambda_\infty)-\psi(-\lambda_\infty)\;,
\end{align}
to zeroth order in $\lambda_\infty$ for $c_\phi$ and $c_\psi$ in terms of $L_\phi$ and $L_\psi$. An important point about AdS$_3$ solutions, which we state here to contrast with the warped AdS$_3$ solutions of later sections, is that the ``non-radial" coordinates (in this case $\phi$ and $\psi$) asymptote to constant values as the affine parameter diverges. In other words, one can safely take the limit $\lambda_\infty\rightarrow \pm \infty$ in either $L_\phi\,$ or $L_\psi$. The solutions to the geodesic equations of motion have two primary branches, which we call the ``cosh-like" and ``sinh-like" branches. We will consider the ``cosh-like" branch, defined by $x(\lambda_\infty)=-x(-\lambda_\infty)$, although the ``sinh-like" branch, defined by $x(\lambda_\infty)=x(-\lambda_\infty)$, can be handled analogously (see Appendix \ref{poinadssols} for details). Using $c=3\ell/2G_N$, we find
\be
S_{\mathrm{EE}}=\f{c}{3}\,\log\;\left(\f{1}{\epsilon}\;\sqrt{L_{\psi}\,\ell\sinh\left(\f{L_{\phi}}{2 \ell}\right)}\right)\,.\label{poinfiber}
\ee
We will comment in the next section on what $\ell$, the curvature scale, is doing in a field-theory formula. Given that the geometry we are considering is simply a coordinate transformation of the usual Poincar\'e patch on AdS$_3$, we could have gotten this answer by performing the appropriate transformations on the usual Poincar\'e patch answer, $(c/3) \log (L/\epsilon)$. This method is easier since the Poincar\'e patch is globally static, allowing us to use the time-independent proposal, and the geodesics are semicircles. To see how such an approach works, see Appendix \ref{appendix}. We will increasingly rely on using such coordinate transformations as we begin warping the spacetime in the later sections.

\subsubsection{Interpretation}\label{feffgraham}
We can suggestively rewrite the answer for the entanglement entropy as
\be
S_{\mathrm{EE}}=\f{c}{3}\,\log\;\f{\sqrt{L_{\psi}\,\ell\sinh\left(\f{L_{\phi}}{2 \ell}\right)}}{\epsilon}=\frac{c}{6}\,\log\,\f{L_\psi}{\epsilon}+\f{c}{6}\,\log\left(\f{\ell}{\epsilon}\sinh\left(\f{L_\phi}{2\ell}\right)\right)\label{heesuggestive}.
\ee
This answer looks like the ground-state answer in the $\psi$ direction and the finite-temperature answer in the $\phi$ direction, with the temperature being set by the curvature scale $\ell$. Recall that $\phi$ and $\psi$ are null coordinates on the conformal boundary, so these correspond to the left- and right-movers.

To investigate the dual state corresponding to this bulk geometry, we can write out the bulk metric near the boundary in the Fefferman-Graham expansion \cite{fgexpansion}:
\begin{align}
  ds^2=\ell^2\left(\f{d\rho^2}{4\rho^2}+h_{ij}(x^i,\rho)dx^i dx^j\right), \qquad h_{ij}(x^i,\rho)=\f{g_{ij}^{(0)}}{\rho}+g_{ij}^{(2)}+\dots\;.
\end{align}
In general, the boundary metric $g_{ij}^{(0)}$ determines the trace and covariant divergence of $g_{ij}^{(2)}$ through the equations of motion near the boundary as
\begin{align}
  \Tr \,g^{(2)}&\equiv g^{(0)}_{ij}g^{(2) ij}=-\f{1}{2}\,R[g^{(0)}_{ij}], \\
  \nabla_i g^{(2)ij}&=\nabla^j \Tr\, g^{(2)},
\end{align}
where the covariant derivative is with respect to the metric $g_{ij}^{(0)}$. The expectation value of the stress-energy tensor is then given by the variation of the renormalized on-shell action with respect to $g^{(0)}_{ij}$  \cite{Balasubramanian:1999re, de Haro:2000xn}, which in two boundary dimensions turns out to be
\begin{align}
\langle T_{ij} \rangle = \f{\ell}{8\pi G}\left(g_{ij}^{(2)}-g_{ij}^{(0)} \Tr \,g^{(2)}\right).
\end{align}
For the usual Poincar\'e patch, we identify $g_{ij}^{(2)}=0$, so we see that $\langle T_{ij} \rangle = 0$. However, for the Poincar\'e fibered coordinates, since $g_{\phi \phi}^{(2)}=1/4$ we have
\begin{align}
\langle T_{\phi \phi} \rangle = \f{\ell}{32\pi G}=\f{c}{48\pi}\label{stressen}
\end{align}
with all other components vanishing (the tracelessness of the stress-energy tensor is preserved since $g_{\phi\phi}^{(0)}=0$). Thus, we are not in the vacuum state of the dual theory and should not have expected to get the universal answer for the vacuum state, which in this case would have been
\begin{align}
S_{\mathrm{EE}}=\f{c}{3}\, \log \,\f{\sqrt{L_{\psi}L_{\phi}}}{\epsilon}
\end{align}
since lengths in the boundary metric are computed with $ds^2=d\phi \,d\psi$. In fact, we do get the vacuum answer for the $\psi$-movers, which agrees with $\langle T_{\psi \psi} \rangle =\langle T_{\psi \phi} \rangle = 0$. The $\phi$-movers are in an excited state, which agrees with $\langle T_{\phi \phi} \rangle\neq 0$. As it should, the bulk diffeomorphism that takes one from Poincar\'e coordinates to Poincar\'e fibered coordinates induces a conformal transformation on the boundary theory, and (\ref{stressen}) is just what one obtains by conformally transforming the vanishing stress-energy tensor from  Poincar\'e coordinates to Poincar\'e fibered coordinates.

Now that we have shown that the modes in the $\psi$ direction are in their ground state and the modes in the $\phi$ direction are excited, the expression for the entanglement entropy is becoming a bit clearer. To make the finite-temperature interpretation more precise, we consider the metric (\ref{fads3}) with compactified fiber coordinate:
\be
  ds^2=\f{1}{4}\left(-\ell^2\frac{d\psi^2}{x^2}+\ell^2\f{dx^2}{x^2}+\left(d\phi+\ell\f{d\psi}{x}\right)^2\right), \qquad \phi\sim\phi+4\pi r_+\,.\label{fads3compact}
\ee
This is precisely the geometry that appears in a Penrose-like near-horizon limit of the extremal BTZ black hole
\be
ds^2=-\f{(r^2-r_+^2)^2}{r^2\ell^2}dt^2+\f{\ell^2r^2}{(r^2-r_+^2)^2}dr^2+r^2\left(d\phi-\f{r_+^2}{\ell r^2}dt\right)^2,\label{btz}
\ee
which has dimensionless $J=M=\f{2r_+^2}{\ell^2}$ and $S=4\pi r_+$ in units where $8G=1$.  Defining left-moving and right-moving energies as
\be
E_L\equiv M-J=0, \qquad E_R\equiv M+J,
\ee
and dimensionless left-moving and right-moving temperatures as
\be
T_L\equiv \ell \;\f{\p E_L}{\p S}=0\,,\qquad T_R\equiv \ell\;\f{\p E_R}{\p S}=\f{r_+}{\pi \ell}\,,
\ee
we see that the state dual to the background (\ref{fads3}) is at zero left-moving temperature and finite right-moving temperature. Though it is at zero Hawking temperature, the statistical degeneracy is explained by the Cardy formula and the nonvanishing right-moving temperature:
\be
S=\f{\pi^2}{3}(c_LT_L+c_RT_R)=\f{\pi^2}{3}\f{3\ell}{2G_N}\f{r_+}{\pi \ell}=4\pi r_+\,,
\ee
which matches the area of the horizon in coordinates (\ref{fads3compact}) or (\ref{btz}). We have used $8G_N=1$ to get to the final expression.

Notice that our answer (\ref{heesuggestive}) applies for the geometry with compact fiber coordinate (\ref{fads3}) as long as we consider small $L_\phi$. With the thermodynamic language developed above, we can define $\tilde{\phi}=\phi/(2r_+)$, $\tilde{\psi}=r_+\psi/\pi\ell=\psi/\beta_R$ and rewrite the second piece in (\ref{heesuggestive}) as
\be
\f{c}{6}\,\log\left(\f{\ell}{\epsilon}\sinh\left(\f{L_\phi}{2\ell}\right)\right)\longrightarrow\f{c}{6}\,\log\left(\f{\beta_R}{\epsilon}\sinh\left(\f{\pi \, L_{\tilde{\phi}}}{\beta_R}\right)\right),\label{temprewrite}
\ee
where $\tilde{\phi}\sim\tilde{\phi}+2\pi$ and the first term in \eqref{heesuggestive} remains unchanged. The UV-IR relation is fixed to match onto the ground state answer in the limit of small $L_{\tilde{\phi}}$. So we have seen that the entanglement entropy answer for the geometry with a compact fiber coordinate reflects the fact that the right-movers are at finite right-moving temperature.

We pause for a moment to connect to an existing result in the literature, which is the calculation of entanglement entropy in the state dual to the rotating BTZ black hole \cite{Hubeny:2007xt}. Taking the extremal limit of their result, $\beta_R\rightarrow \infty$, one finds
\be
S_{\mathrm{EE}}=\frac{c}{6}\,\log\,\f{L}{\epsilon}+\f{c}{6}\,\log\left(\f{\beta_R}{\epsilon}\sinh\left(\f{\pi L}{\beta_R}\right)\right)
\ee
for purely spatial separation on the boundary. This is precisely our answer with $L_{\tilde{\phi}}=L_{\tilde{\psi}}=L$. It seems that the IR limit we have taken to get to the geometry \eqref{fads3compact} has retained the entangling properties of the dual state.

Now we would like to take the limit where the geometry decompactifies, i.e. $r_+/\ell\rightarrow \infty$, since this allows us to recover our original geometry (\ref{fads3}). Notice that in this limit, we are going from having two scales, $\ell$ and $r_+$, to just one scale $\ell$. Thus, all dimensionful parameters must be measured relative to $\ell$.\footnote{The role of the lattice spacing $\epsilon$ will not be important for this argument.} Looking at the left-hand-side of (\ref{temprewrite}), we see that this means that the argument of the $\sinh$ must remain fixed in this limit, since we want to keep $L_\phi$ (in units of $\ell$) fixed. Expressed in terms of the CFT quantities on the right-hand-side of (\ref{temprewrite}), we are taking $\beta_R$ small with $L_{\tilde{\phi}}/\beta_R$ fixed. We therefore retain the interpretation of the right-movers being at finite temperature in the decompactification limit. With compact fiber coordinate, the expression ($\ref{heesuggestive}$) can be understood in relation to the DLCQ limit which freezes the $\psi$-movers to their ground state \cite{Balasubramanian:2009bg}.

To aid with understanding taking arbitrary spacelike slices, we note here that the expression for the length of an extremal geodesic in the rotating BTZ background for arbitrary spacelike separation on the boundary can be written:
\begin{align}
S_{EE}=\f{c}{6}&\log \left[\f{\beta_L\beta_R}{\pi^2\epsilon^2}\sinh \left(\f{\pi\Delta x_L}{\beta_L}\right)\sinh\left(\f{\pi\Delta x_R}{\beta_R}\right)\right]\\
=\f{c}{6}&\log \left[\f{\beta_L}{\pi\epsilon}\sinh\left(\f{\pi\Delta x_L}{\beta_L}\right)\right]+\f{c}{6} \log \left[\f{\beta_R}{\pi\epsilon}\sinh\left(\f{\pi\Delta x_R}{\beta_R}\right)\right]\label{btzsectors}
\end{align}
for $x_L=\phi+t$ and $x_R=\phi-t$. Again, the contribution to the entanglement entropy splits up into distinct contributions from the left- and right-moving sectors. This is analogous to how the contribution to the thermodynamic entropy splits into left- and right-moving sectors in the Cardy formula.

\subsection{Global fibered $\mathrm{AdS}_3$}

The global fibered AdS$_3$ metric is obtained by setting $a=1$ in \eqref{globalfibered} to obtain
\begin{align}
    ds^2 = \frac{\ell^2}{4}\left(-(1+r^2)\,d\tau^2 + \frac{dr^2}{1+r^2}+(du + r\,d\tau)^2\right).\label{globalfiberedads}
\end{align}
All coordinates are dimensionless while $\ell$ has dimensions of length. The coordinates $u$ and $\tau$ become null near the part of the boundary reached by $r\rightarrow \pm\infty$, which is the region to which we shall restrict our attention; see Figure \ref{ads3cylinder} for the precise parameterization of the boundary cylinder in these coordinates. One can write conservation equations for affinely parameterized geodesics just like in the Poincar\'e fibered case.  In this case, we label the conserved quantities corresponding to translations in $\tau$ and $u$ by $c_\tau$ and $c_u$ respectively, while $c_v = \dot x^\mu \dot x_\mu$.  After some manipulation, the conservation equations can be written as follows:
\begin{align}
    \dot r^2
    &= \left(\frac{c_v}{(\ell/2)^2}\right)r^2 -\left(\frac{2c_uc_\tau}{(\ell/2)^4}\right) r-\frac{c_u^2-c_\tau^2-(\ell/2)^2c_v}{(\ell/2)^4}\,, \label{reqads}\\
    \dot \tau
    &= \frac{c_u}{(\ell/2)^2}\frac{r}{r^2+1}  -\frac{c_\tau}{(\ell/2)^2}\frac{1}{r^2+1}\,, \label{taueqads}\\
    \dot u
    &= \frac{c_\tau}{(\ell/2)^2}\frac{r}{r^2+1} + \frac{c_u}{(\ell/2)^2}\frac{1}{r^2+1}\,. \label{ueqads}
\end{align}
Imposing the condition $c_v>0$ ensures that the geodesics determined by these equations are spacelike.  For equation \eqref{reqads}, there are ``cosh-like" and ``sinh-like" solution branches according to whether $c_u^2-c_v(\ell/2)^2>0$ and $c_u^2-c_v (\ell/2)^2<0$, respectively. The solutions are presented in Appendix \ref{globaladssols}.

We now wish to calculate the leading divergent piece of the length of these geodesics. The approach is identical to the previous section, so we will not repeat the details here. Using the UV-IR relation $r_\infty\sim \epsilon^{-2}$ for dimensionless cutoff $\epsilon$, we find
\begin{align}
    \lambda_\infty\approx\frac{(\ell/2)}{\sqrt{c_v}}
    \log\left(\frac{c_v (\ell/2)^2}{\sqrt{(c_u^2-c_v (\ell/2)^2)(c_\tau^2+c_v (\ell/2)^2)}}\frac{1}{\epsilon^2}\right).
\end{align}
We can now trade in the conserved quantities $c_u$ and $c_\tau$ for coordinate separations $L_u$ and $L_\tau$ on the boundary and recover
\begin{align}
    S_{\mathrm{EE}}= \frac{c}{3}\,\log\left(\f{1}{\epsilon}\;\sqrt{\sin\left(\frac{L_\tau}{2}\right)\sinh\left(\frac{L_u}{2}\right)}
  \right).\label{globalfiber}
\end{align}
We will stick to $L_\tau<2\pi$ on the boundary to maintain spacelike separation between the two endpoints ($\tau$ is a null coordinate that winds up the cylinder). Just as in the Poincar\'e fibered case \eqref{poinfiber} we see that the $u$-moving sector seems to be at finite temperature, with the temperature scale set by $\ell$ (recall that our coordinate $u$ is dimensionless), while the $\tau$-moving sector is in its ground state. The appearance of the sine function is simply from the compact $U(1)$ of the global AdS$_3$ cylinder. One can perform a Fefferman-Graham analysis by repeating the steps of Section \ref{feffgraham}, but the details are the same and we omit them here.

The result for the ``sinh-like" branch is similar:
\be
  S_{\mathrm{EE}}= \frac{c}{3}\,\log\left(\f{1}{\epsilon}\;\sqrt{\cos\left(\frac{L_\tau}{2}\right)\cosh\left(\frac{L_u}{2}\right)}\right).\label{sinhbranch}
\ee
Notice that the length remains well-defined when $L_u\rightarrow 0$ and $L_\tau\rightarrow 0$, as it should since the geodesic is going through the bulk from $r=-\infty$ to $r=\infty$ in this limit. We mention this branch due to its relevance to the metric (\ref{globalfibered}) with compact fiber coordinate. This is the self-dual orbifold considered first in \cite{Coussaert:1994tu} and studied extensively in \cite{Balasubramanian:2009bg, Balasubramanian:2003kq}. The geometry is locally AdS$_3$ and has an AdS$_2$ factor, but a compact fiber coordinate causes the two boundaries at $r=+ \infty$ and $r=-\infty$ to become disconnected, though they are \emph{causally connected} through the bulk. The entanglement  between the asymptotic boundaries was computed via a reduction to AdS$_2$/CFT$_1$ in \cite{Azeyanagi:2007bj}. Our answer can be used to compute quantities like the holographic thermo-mutual information (HTMI) in these horizon-less backgrounds, as defined in  \cite{Morrison:2012iz}, directly in AdS$_3$.

\section{Spacelike $\mathrm{WAdS}_3$}\label{wads3}
We have seen in the previous sections how to apply the covariant HRT proposal to locally AdS$_3$ spacetimes written as a real-line fibration over AdS$_2$. The results agree with the universal CFT$_2$ answers for a state at zero left-moving temperature and finite right-moving temperature. We now move on to the case of nontrivial warping. We will set up the problem with general warping parameter $a\neq 1$ and only specify our peturbative expansion about AdS$_3$ with $a=1+\delta$ at a later point in our analysis.

\subsection{Global coordinates analysis}

We consider the metric \eqref{globalfibered}, and we determine the affinely parameterized geodesics $x^\mu(\lambda) = (\tau(\lambda), u(\lambda), r(\lambda))$ in this geometry.  The metric has Killing vectors $\partial_\tau$ and $\partial_u$ corresponding to translations in $\tau$ and $u$, and they yield conserved quantities $c_\tau = \dot x\cdot\partial_\tau$ and $c_u = \dot x\cdot\partial_u$, respectively.  Since we consider affinely parameterized geodesics, the square speed $ c_v = \dot x^\mu \dot x_\mu$ along the geodesic is also conserved.  The corresponding conservation equations are
\begin{align}
    c_\tau
    &= (\ell/2)^2 \left(a^2 r \left(r \dot{\tau}+\dot{u}\right)- \left(r^2+1\right) \dot{\tau}\right), \label{ctau}\\
    c_u
    &= (\ell/2)^2a^2 \left(r \dot{\tau}+\dot{u}\right), \label{cr} \\
    c_v
    &= \frac{(\ell/2)^2 \left[- \left(r^2+1\right)^2 \dot{\tau}^2+a^2 \left(r^2+1\right) \left(r \dot{\tau}+\dot{u}\right)^2+\dot{r}^2\right]}{r^2+1}\;. \label{cE}
\end{align}
To solve these equations, it helps to manipulate them into the following form:
\begin{align}
    \dot{r}^2
    &= -\left(\frac{c_u^2(1 - a^2)-(\ell/2)^2a^2c_v}{(\ell/2)^4a^2}\right)r^2 -\left(\frac{2c_uc_\tau}{(\ell/2)^4}\right) r-\frac{c_u^2-a^2c_\tau^2-(\ell/2)^2a^2c_v}{(\ell/2)^4a^2}\;, \label{req}\\
    \dot{\tau}
    &= \frac{c_u}{(\ell/2)^2}\frac{r}{r^2+1}  -\frac{c_\tau}{(\ell/2)^2}\frac{1}{r^2+1} \;,\label{taueq}\\
    \dot{u}
    &= \frac{c_\tau}{(\ell/2)^2}\frac{r}{r^2+1} + \frac{c_u}{(\ell/2)^2a^2}\frac{1}{r^2+1}+\frac{c_u}{(\ell/2)^2}\left(\frac{1-a^2}{a^2}\right)\frac{r^2}{r^2+1}\;. \label{ueq}
\end{align}
Equation \eqref{req} is now a decoupled, separable differential equation that can be integrated to determine $r(\lambda)$.  The solution to \eqref{req} can then be plugged into equations \eqref{taueq} and \eqref{ueq}, which can be integrated to obtain $\tau(\lambda)$ and $u(\lambda)$, respectively.  Notice also that setting $a=1$ in these equations gives the system of equations \eqref{reqads}, \eqref{taueqads}, and  \eqref{ueqads}. The equations become $a$-independent in the limit $c_u=0$, though such a limit does not seem particularly useful for understanding warped AdS$_3$; see Appendix \ref{cueq0}. The general solutions to these equations can be found in Appendix \ref{globalwadssols}. We simply note here that the solution for $u(\lambda)$ has a piece that grows linearly with $\lambda$, unlike in the AdS$_3$ case. This means that the relation between $c_u$ and $L_u$ will necessarily involve $\lambda_\infty$. This complicates the analysis, as we shall see shortly.

Let us focus on the ``cosh-like" branch with $0<a<2$ and fix $c_v=1$. This includes the squashed and stretched cases. In our approach, we first write the length of the geodesic in terms of the conserved quantities and the cutoff in the holographic coordinate $r$:
\be
\lambda_{\infty}= \f{1}{\sqrt{c_1}}\cosh^{-1}\left[\f{-c_2+2c_1r_\infty}{\sqrt{c_2^2+4c_1c_3}}\right],\label{lengthform}
\ee
where we have used the definitions in (\ref{constants}) and require $c_1>0$. This expression holds for general warping $a$ as well as for the AdS$_3$ case of $a=1$ (the $a$-dependence is buried in $c_1$ and $c_3$). Taking $c_1r_\infty\gg c_2$ and restoring the original constants of motion $c_u$ and $c_\tau$ gives\footnote{One cannot in general be so cavalier in taking $c_1r_\infty \gg c_2$ without any restrictions on $L_u$, since $c_2$ depends on $c_u$, which depends on $\lambda_\infty$. In our case, however, this can be consistently realized by taking $r_\infty\gg 1$.}
\be
\lambda_\infty\approx\f{\log\left[\f{c_1r_\infty}{\sqrt{c_2^2+4c_1c_3}}\right]}{\sqrt{c_1}} =\f{\log\left[r_\infty\f{a^2(1+c_u^2)-c_u^2}{\sqrt{(-a^2+c_u^2)(a^2(1+c_\tau^2+c_u^2)-c_u^2)}}\right]}{\sqrt{1+(1-1/a^2)c_u^2}}\;,\label{ansconst}
\ee
where we have set $c_v=\ell/2=1$. Although the geodesic equations for $r(\lambda)$, $\tau(\lambda)$, and $u(\lambda)$ are soluble, to write the answer for the length in terms of coordinate separations on an asymptotic boundary (instead of in terms of conserved quantities as done above) there remains the task of inverting limits of those solutions to obtain the conserved quantities $c_u$ and $c_\tau$ in terms of separations on the boundary $L_u$ and $L_\tau$. The equation for the $\tau$ coordinate is simply generalized from the AdS case:
\be
c_\tau=\sqrt{c_1}\cot\left(\frac{L_\tau}{2}\right)=\sqrt{\f{c_u^2(a^2-1)+a^2}{a^2}}\cot\left(\frac{L_\tau}{2}\right),
\ee
which holds as long as $L_\tau<\pi$. The new feature in these spacetimes, which is different from asymptotically AdS spacetimes, is that the relation between $c_u$ and $L_u$ involves $\lambda_{\infty}$:
\be
2\left(-1+\f{1}{a^2}\right)c_u\lambda_{\infty}+\log\left(\f{c_u+\sqrt{1+c_u^2-\f{c_u^2}{a^2}}}{c_u-\sqrt{1+c_u^2-\f{c_u^2}{a^2}}}\right)=L_u\;.\label{trouble}
\ee
In other words, one cannot keep both $c_u$ and $L_u$ fixed as the cutoff is scaled large. This follows directly from the linear divergence of $u(\lambda)$ with $\lambda$, as would occur in an AdS$_2\times \mathbb{R}$ background. One could at this point try to proceed by solving for $c_u$ in terms of $L_u$ and $\lambda_\infty$ and plug $c_u$ and $c_\tau$ into \eqref{ansconst}. This would then be an equation for $\lambda_\infty$ that can be solved to determine the length. Unfortunately, such an approach has two obstacles, one conceptual and one technical. The conceptual obstacle is that this would correspond to fully applying the HRT proposal in an asymptotically warped AdS$_3$ spacetime, and it is unclear whether such a prescription makes sense. The technical obstacle (at least in this approach) is that \eqref{trouble} is a transcendental equation for $c_u$. In the case of AdS$_2\times\mathbb{R}$, which can be realized as the $a\rightarrow 0$ limit of warped AdS$_3$, the left-hand-side of the analog of \eqref{trouble} has only the piece linear in $\lambda_\infty$ and such a method can be carried out.

In the next section, we will show that setting up a perturbative expansion about the AdS$_3$ point by considering warping parameter $a=1+\delta$ will allow us to solve this equation order-by-order in $\delta$.

\section{Perturbative entanglement entropy}\label{pertsec}

Given that a nonperturbative application of the HRT prescription to asymptotically warped AdS$_3$ spacetimes is suspect, here we will try to infinitesimally perturb around the AdS$_3$ point and use the AdS/CFT dictionary, which presumably contains as one of its entries the HRT prescription. Deep in the IR, the geometry (\ref{globalfibered}) is close to AdS$_3$, and it is only in the UV that the nontrivial warping parameter begins to destroy the asymptotics. If we cut off our spacetime before this happens, then we are at low enough energies where our analysis will be on firmer ground. Viewed in this way, we have a conformal field theory which we perturb by an infinitesimal, irrelevant operator. Holographic renormalization can then be understood perturbatively in this infinitesimal source \cite{vanRees:2011fr, vanRees:2011ir}. We will find that in a certain limit we can sum the perturbative expansion to \emph{all} orders. The resulting answer takes the precise form of a two-dimensional CFT and reproduces the warping-dependent central charge and left- and right-moving temperatures postulated previously in the literature.

\subsection{Perturbative expansion}
We imagine that the warping parameter is close to $1$, i.e. $a=1+\delta$ for $|\delta| \ll 1$. It is in this sense which we expand about the AdS$_3$ point $a=1$. Such a perturbative expansion will help us solve (\ref{trouble}) for $c_u$ order-by-order in $\delta$. The solutions below follow a simple pattern at each order, and though we list the general formulae for arbitrary order, we have technically only checked that they are true to tenth order. Expanding
\be
c_u=c_{u,0}+\delta \,c_{u,1}+\delta^2 c_{u,2}+\cdots,
\ee
we solve (\ref{trouble}) to get
\begin{align}
  c_{u,0}&=(\ell/2)\,\textrm{coth }\f{L_u}{2}\;, \\
  c_{u,1}&=\f{1}{2}((\ell/2)-4\lambda_\infty+(\ell/2)\cosh{L_u})\;\textrm{coth }\f{L_u}{2}\;\textrm{csch}^2\f{L_u}{2}\;,\\
  c_{u,n}&=\left( \sum_{j=0}^{n-1}\sum_{i=0}^{n}\lambda_\infty^i (\ell/2)^{n-i}k_{ij}^{(n)}\cosh (jL_u) \right)\textrm{coth } \f{L_u}{2}\; \textrm{csch}^{2n} \f{L_u}{2}\;;\qquad n>1\,,\label{cuhigh}
\end{align}
where the $k_{ij}^{(n)}$ are calculable $n$- and $\lambda_\infty$-dependent constants. We require $|\delta^n\,c_{u,n}|\ll |\delta^{n-1}c_{u,n-1}|$ to assure convergence of our perturbative expansion. This can be satisfied by taking
\be
L_u\gtrsim 1\,,\qquad |\lambda_\infty\, \delta| \ll 1\,.
\ee
$L_u$ is being measured in units of $\ell$. The latter condition ensures that we stay in an AdS$_3$-like part of the geometry and not get into the WAdS$_3$ asymptotic. Notice that from the point of view of perturbing about AdS$_3$, this is an eminently sensible condition; regardless of how small one takes $\delta$, the geometry looks wildly different from AdS$_3$ for sufficiently large $\lambda_\infty$, so we need to constrain their product. Incidentally,  the curvature invariants of WAdS$_3$ are all finite and continuously connected to the AdS$_3$ case $a=1$, so they are not a good way to classify where to cut off the spacetime for a well-defined perturbation theory. When computing the length, we keep only the leading divergent piece (in $r_\infty$) at each order. This gives the following result for the entanglement entropy:
\begin{align}
  S_{\mathrm{EE}}=\,&\f{\ell}{4G_N}\left[\left(1+\delta\, \textrm{coth}^2\f{L_u}{2}\right)\log\left(r_\infty \sin\,\f{L_\tau}{2}\sinh\,\f{L_u}{2}\right)\right]+\nonumber
  \\ &\f{\ell}{4G_N}\sum_{i=2}^\infty \delta^i\,(-1)^{i+1}\textrm{coth}^2\f{L_u}{2}\textrm{csch}^{2(i-1)}\f{L_u}{2}\left[\log\left(r_\infty \sin\,\f{L_\tau}{2}\sinh\,\f{L_u}{2}\right)\right]^i\left(\sum_{j=0}^{i-2}c_{ij}\cosh(jL_u)\right).\label{everything}
\end{align}
The constants $c_{ij}$ are all positive. Notice that the zeroth order piece is precisely the answer for AdS$_3$ given in (\ref{globalfiber}), as it should be. Unfortunately, the series does not seem simply summable unless we take the scaling limit $L_u\gg 1$, in which case we use our knowledge of the $c_{ij}$ and sum the series to get
\begin{align}
  S_{\mathrm{EE}}=\,&\f{\ell}{4G_N}\left[\left(1+\delta\right)\log\left(r_\infty \sin\,\f{L_\tau}{2}\sinh\,\f{L_u}{2}\right)\right]-\nonumber \\
  &\f{\ell}{4G_N}\,e^{-L_u}\left[-1+4\delta \log\left(r_\infty \sin\,\f{L_\tau}{2}\sinh\,\f{L_u}{2}\right)+e^{-4\delta \log\left(r_\infty \sin\,\f{L_\tau}{2}\sinh\,\f{L_u}{2}\right)}\right]
\end{align}
to leading order in $L_u$.  Notice that the first two terms in the second line are suppressed by a factor of $e^{-L_u}$ relative to the first line and can safely be dropped. Up to an overall constant, the last term in the second line can be written as
\begin{align}
  \left(\sin\,\f{L_\tau}{2}\right)^{-4\delta}\Big(r_\infty ^{-4\delta} e^{-L_u(1+2\delta)}\Big).
\end{align}
For $\delta>0$, this is suppressed relative to the first line without further qualification and can be dropped. For $\delta < -1/2$, this term grows with $L_u$ and cannot be neglected.  However, for $-1/2<\delta < 0$, there is a competition between the factor containing $r_\infty\sim e^{\lambda_\infty}\gg 1$ and the factor containing $L_u\gg 1$.  In this regime, for a given $\delta$ and $r_\infty$, one simply needs to choose $L_u$ sufficiently large ($L_u(1+2\delta)\gg -4\delta \lambda_\infty$) such that the resulting expression is dominated by the expression in the first line.

By combining these observations, we find that if $\delta>-1/2$, then the leading behavior of the entanglement entropy in the large-$L_u$ regime is
\be
  S_{\mathrm{EE}}=\f{\ell}{2G_N}\left(1+\delta\right)\log\left(\f{1}{\epsilon}\;\sqrt{\sin\,\f{L_\tau}{2}\exp\left(\f{L_u}{2}\right)}\right), \label{weeir}
\ee
where we have used the UV-IR relation $r_\infty\sim 1/\epsilon^2$. Since we are sourcing an infinitesimal irrelevant operator and computing perturbatively, the UV-IR relation used should remain that of AdS/CFT. We have also replaced the hyperbolic sine function with an exponential function, since corrections are subleading in our expansion in $e^{-L_u}$. As usual, numerical factors are absorbed into a redefinition of the cutoff $\epsilon$. 

We see that for $a=1+\delta$, the perturbative expansion in the large-$L_u$ gives simply the two-dimensional CFT answer of (\ref{globalfiber}) upon identifying the coefficient of the logarithm with $c/3$:
\be
  c_L=c_R=\f{3\ell}{2G_N}(1+\delta)\,.\label{ourseries}
\ee
The equality of $c_L$ and $c_R$ is due to a lack of diffeomorphism anomaly, since we are working in Einstein gravity. These are precisely the central charges of \cite{Anninos:2008qb}, conjectured  by demanding consistency with the Cardy formula (we will reproduce this check in Section \ref{nonperturbative}).\footnote{To facilitate comparison with the notation of \cite{Anninos:2008qb}, one should take $a^2\rightarrow \beta^2$ and $\ell^2\rightarrow (4-\beta^2)\ell^2/3$. Notice that as $\beta^2\rightarrow 4$, which is the limit in which the central charge of \cite{Anninos:2008qb} vanishes, there is an infinite rescaling that allows our central charge to remain finite.} One of these central charges has been produced through an asymptotic symmetry group analysis \cite{Compere:2007in}. Identifying the functional form of $S_{EE}$ with the AdS$_3$ result (\ref{globalfiber}) allows us to conclude that the dual state lives on a cylinder charted by null coordinates $\tau$ and $u$.

It is important to keep in mind that the entanglement entropy computed in \eqref{weeir} is understood as  an expansion to zeroth order in $e^{-L_u}$ but to \emph{all} orders in $\delta$. This approach can in principle be extended to lower orders in $L_u$, and the appearance of the logarithmic term in our general formulae suggests that the answer will remain roughly in the form of the CFT$_2$ answer, except the logarithm will have an $L_u$-dependent prefactor. This is consistent with the existence of a single Virasoro algebra, since it seems the answer only picks up additional $u$-dependence while keeping the $\tau$-dependence the same. Our result at leading order in $e^{-L_u}$ seems to suggest that warped CFTs behave like ordinary CFTs in the IR, for large $L_u$. The IR restriction is due to cutting off our spacetime deep in the bulk and is independent of the large $L_u$ restriction. The similarity to CFT$_2$ jibes well with the fact that the deep interior of the WAdS$_3$ geometry is AdS$_3$-like. We will discuss the physical meaning of large $L_u$ in Section \ref{largefiber}. We will also go beyond the small warping limit in Section \ref{nonperturbative} by arguing that warped CFTs are CFT-like generally, as long as one takes an infrared limit and studies large $L_u$.

Performing the same perturbative expansion in the case of Poincar\'e coordinates would give a result that can be obtained simply by coordinate transforming our current answer as in Appendix \ref{fiberedcoords}, and it is given by
\be
  S_{\mathrm{EE}}=\f{\ell}{2G_N}\left(1+\delta\right)\log\left(\f{1}{\epsilon}\;\sqrt{ L_\psi \ell\,\exp\left(\f{L_\phi}{2\ell}\right)}\right).\label{eepoinwads}
\ee
This is again the appropriate answer at large $L_\phi$ for a two-dimensional CFT, now on the Minkowski plane charted by null coordinates $\phi$ and $\psi$, as presented in (\ref{poinfiber}). Taking the fiber coordinate $L_u$ large in the global coordinate system corresponds to taking the fiber coordinate $L_\phi$ large in Poincar\'e coordinates. In the case of Poincar\'e coordinates, however, we can simultaneously take $L_\psi$ large if we want to consider a particular time slice $L_\phi=L_\psi$.

Due to the convergence of the perturbative expansion for any warping parameter $a>1/2$ in the large fiber-coordinate regime, we conjecture that the nonperturbative answer for the entanglement entropy for a state at zero left-moving temperature and finite right-moving temperature, in the large fiber-coordinate regime, is given by \eqref{eepoinwads} for a state on the plane or \eqref{weeir} for a state on the cylinder. We will expound on this conjecture in Section \ref{nonperturbative} after providing some more evidence for our approach. However, we will henceforth use the nonperturbative parameter $a$ in our formulae.

\subsection{Finite temperature}\label{finitetemp}

In the limit of large separation in the fiber coordinate, we can match our results with those of two-dimensional CFT even at finite temperature. Since black holes in warped AdS$_3$ are given by discrete quotients of the vacuum spacetime, they are locally warped AdS$_3$ \cite{Anninos:2008fx}. This is analogous to BTZ black holes in AdS$_3$. Due to the local equivalence, we can exhibit local coordinate transformations that take us from the geometry with a black hole to the geometry without a black hole. We will stick to the stretched case $a>1$ to avoid closed timelike curves. The metric for the warped BTZ black hole is given by
\begin{align}
\frac{ds^2}{\ell^2}&=\f{3dt^2}{4-a^2}+\frac{dr^2}{4(r-r_+)(r-r_-)}+\frac{6\sqrt{3}}{(4-a^2)^{3/2}}\left(a r-\sqrt{r_+ r_-}\right)dt d\theta \nonumber \\&+\frac{9r}{(4-a^2)^2}\left((a^2-1)r+r_++r_--2a\sqrt{r_+r_-}\right)d\theta^2\;.\label{warpedBTZ}
\end{align}
We will restrict to the stretched case $a>1$ to avoid the presence of closed timelike curves at large radial coordinate. The answer for the entanglement entropy in a warped BTZ background can be reproduced by coordinate transforming our previous answer. The coordinate transformations can be found in Section 5 of \cite{Anninos:2008fx}. Performing such a transformation to  \eqref{weeir}, we find
\be 
  S_{\mathrm{EE}}=\frac{\ell a}{G_N}\log\left(\f{r_+-r_-}{\epsilon^2}\,\exp\left(\sqrt{\frac{3}{a^2(4-a^2)}}\,\Delta t+\f{\pi\Delta\theta}{\beta_L}\right) \sinh\f{\pi\Delta\theta}{\beta_R}\right), \label{wbtz}
\ee
with dimensionless temperatures
\begin{align}
\beta_L^{-1}&=T_L=\frac{3}{2\pi (4-a^2)}\left(r_++r_--\frac{2}{a}\sqrt{r_+r_-}\right), \label{betaL}\\
\beta_R^{-1}&= T_R=\frac{3(r_+-r_-)}{2\pi (4-a^2)}\;. \label{betaR}
\end{align}
Due to the compactification of $\theta$, there can exist many spacelike geodesics in this geometry, distinguished by their winding number and directionality. The expression $\Delta\theta$ refers to the separation in a noncompact $\theta$, i.e. without modding by $2\pi$. We can ignore the global topology by considering $\Delta\theta\ll 2\pi$. This is consistent with the large-$L_u$ limit taken in the previous section, since that limit can be accomodated by taking $\Delta t$ large. Adding winding will only increase the length of the geodesic,\footnote{Note that this would be more subtle if we considered the squashed case $a<1$, since for large enough $r$ winding in $\theta$ corresponds to a timelike direction and can \emph{decrease} the length of the geodesic.} so we see that our answer is valid in the regime considered. 

In the case of AdS$_3$ with $a=1$, the coordinates are such that one picks a constant-time slice by requiring $\Delta t=0$. It is important to note that this case corresponds to the BTZ black hole in a rotating coordinate system, and our answer for $a=1$ is the universal CFT$_2$ answer for such a dual state. Since we are using a rotating coordinate system, it is not necessary that the functional form of our answer precisely match the form of \eqref{btzsectors}. The parameters $\beta_L$ and $\beta_R$ give the inverse left-moving and right-moving temperatures of the BTZ black hole in this frame, and we see that this match extends to the warped BTZ case as well; the dimensionful temperatures (\ref{betaL}) and (\ref{betaR}) match precisely with those of \cite{Anninos:2008fx}. In Section \ref{nonperturbative} we will show that these temperatures, combined with the central charge \eqref{ourseries}, satisfy the Cardy formula. Finally, implementing an appropriate homology constraint suffices to reproduce the thermodynamic black hole entropy in the limit where we consider the entire boundary density matrix without tracing out any degrees of freedom.

\subsection{A vacuum state proposal}\label{vacuum}

We have produced the universal CFT results for states dual to spacelike warped AdS$_3$ and the warped BTZ black hole. However, as our formulae in the previous sections illustrate, none of these states can be considered the vacuum state. The proposal in \cite{Detournay:2012pc} is that the timelike warped AdS$_3$ geometry is a suitable candidate for the vacuum state in both topologically massive gravity and a specific string theory example that reduces to Einstein gravity plus matter. The proposed vacuum geometry (which is in fact G\"{o}del space) can be written as
\be
  \f{ds^2}{\ell^2}=-\f{3dt^2}{4-a^2}+\f{3dr^2}{4r(4-a^2+3r)}-\frac{6ar\sqrt{3}}{(4-a^2)^{3/2}} dt d\theta +\frac{3r(4-a^2-3r(a^2-1))}{(4-a^2)^2} d\theta^2\,,
\ee
where $\theta$ is a compact coordinate with $\theta\sim\theta+2\pi$. For $a^2 > 1$ this geometry has closed timelike curves for $r > (4-a^2)/3(a^2-1)$ (see \cite{Levi:2009az,Banados:2005da} for a discussion). In our perturbative approach, we can take $r_\infty \delta\ll 1$, which is sufficient to excise the region with closed timelike curves. Notice that we can get to this geometry by taking the warped BTZ black hole \eqref{warpedBTZ} and performing the identifications
\be
r_+=0, \qquad r_-=\frac{a^2-4}{3},\qquad t\rightarrow i t, \qquad \theta\rightarrow i \theta\;.
\ee
If we replace the exponential function in \eqref{wbtz} with a hyperbolic sine (i.e. start with a precise match to 2D CFT instead of  a match only at large fiber coordinate), then we can perform these identifications on the entanglement entropy result to get, for $\Delta t=0$,
\be
S_{\mathrm{EE}}=\frac{\ell a}{2G_N}\log \left(\f{\sin(L_\theta/2)}{\epsilon}\right).
\ee
We see that this is the ground state answer for a two-dimensional CFT on a cylinder with a compact spatial coordinate $\theta$. Unfortunately, this is merely illustrative because it runs afoul of the requirement of large fiber-coordinate separation. The correct way to get the answer in our framework is to keep the entire expression \eqref{everything} and perform the identifications necessary to get to timelike warped AdS. From here, there does not appear to be a sensible regime in which the series can be summed and reduces to a two-dimensional CFT answer.

\subsection{The geometric meaning of large fiber-coordinate separation}\label{largefiber}

We now discuss the meaning of the limit of large fiber-coordinate separation $L_\phi$ on a field theory calculation of entanglement entropy. Note that the limit is not necessarily a restriction on the spatial size, $(L_\phi L_{\psi} )^{1/2}$, for which our result holds.\footnote{Here we are referring to spacelike warped AdS$_3$ in Poincar\'e coordinates \eqref{poinfibered}, where the fiber coordinate is denoted by $\phi$, and there is no restriction on the separation in the other coordinate $\psi$. The dual state is on the Minkowski plane and has finite right-moving temperature.} The different spatial sizes lie on spacelike slices boosted with respect to one another. For example, large spatial sizes are accommodated by taking $L_\psi\sim L_\phi$, which results in a ``mostly spacelike" slice, whereas small spatial sizes are accommodated by taking $L_\psi$ small, which makes the slice more null. Nevertheless, imposing large $L_\phi$ without constraining the system size \emph{does} impose a physical restriction on the reduced density matrix. Unlike the case of the vacuum state on the Minkowski plane, there is no Lorentz symmetry relating the different observers on their different spacelike slices. In the case of spacelike warped AdS$_3$, our result for $S_{EE}$ exhibits that there is a finite right-moving temperature turned on, which breaks Lorentz invariance. In the case of warped BTZ black holes, there is also a finite left-moving temperature. Thus, Lorentz transformations connecting different observers act nontrivially and lead to a different reduced density matrix. On the other hand, for the vacuum state on the plane the answer can be boosted and replaced with the invariant Minkowskian interval, as shown in \eqref{boost}. The entanglement entropy in this case is only sensitive to the length of the spatial interval and not the orientation of the spatial slice, whereas when Lorentz invariance is broken it is sensitive to both.

\subsection{Nonperturbative conjecture}\label{nonperturbative}
We have seen that the perturbative series we constructed converges for $a>1/2$ in the large fiber-coordinate regime. Recall that the physically relevant range is $a\in [0,2)$, so we fail to capture part of the parameter space. The region $a\in [0,1/2)$ includes the interesting case of AdS$_2\times \mathbb{R}$, which can be reached by taking $a\rightarrow 0$ and rescaling the fiber coordinate $u\rightarrow u/a$ in \eqref{globalfibered}.

The convergence of our series seems to suggest that our results for the entanglement entropy hold nonperturbatively in the warping. We conjecture this to be true. This claim requires a UV-IR relation of the form $r_\infty \sim 1/\epsilon^2$ to hold nonperturbatively. In our perturbative approach we could make use of this UV-IR relation since we were working in the context of AdS/CFT, where it is known to be true. Extending the requirement into the nonperturbative regime is a natural choice. With it, we claim that our perturbative expansion is sufficient to capture the nonperturbative dynamics entering into the entanglement entropy.

A nontrivial check on this nonperturbative proposal is the Cardy formula. Our answers for $S_{EE}$ allow us to read off left-moving and right-moving temperatures and the central charge. We now claim that all these results hold nonperturbatively. The central charge is given universally as
\be
c_L=c_R=\f{3\ell a}{2G_N}\,.
\ee
For the warped BTZ black hole, our proposal allows us to identify the left-moving and right-moving temperatures as \eqref{betaL} and \eqref{betaR} nonperturbatively in $a$. These temperatures and the central charge reproduce the entropy of the warped BTZ black hole through the Cardy formula:
\begin{align}
S=\f{A}{4G_N}=\left(\f{3\pi\ell}{2G_N(4-a^2)}\,(a r_+-\sqrt{r_+r_-})\right)=\f{\pi^2}{3}(c_LT_L+c_RT_R).
\end{align}

\section{Summary and outlook}\label{summsec}
We have taken the first steps toward understanding holographic entanglement entropy in the context of asymptotically warped AdS$_3$ spacetimes in Einstein gravity. We began by considering AdS$_3$ as a real-line fibration over AdS$_2$, a coordinate system relevant to the study of extremal black holes. The calculation of the entanglement entropy indicated a state at zero left-moving and finite right-moving temperature, as expected.

Deforming the fibration by a nontrivial warp factor leads to the warped AdS$_3$ geometries, appearing in the near-horizon limit of extremal Kerr black holes at constant polar angle. To connect with the HRT proposal in AdS/CFT, we constructed a perturbation theory about the AdS$_3$ point with trivial warping. For $a=1+\delta$, one can compute the length of the necessary geodesic perturbatively in $\delta$ to all orders. The general answer is not particularly illuminating, except in the limit of large separation in the fiber coordinate. Recall that the $U(1)$ isometry originating from translation invariance in the fiber coordinate is what is expected to enhance to an infinite-dimensional $U(1)$ Kac-Moody algebra in the boundary theory. In this limit, the answer takes the universal form predicted by two-dimensional CFT. Interpreting our answer as a CFT answer allows us to read off the purported central charge of the dual theory, which is given by $c=3\ell a/2G_N$. Since we are working in Einstein gravity, there is no diffeomorphism anomaly and $c_L=c_R=c$. Furthermore, heating up the dual state with a warped BTZ black hole in the bulk again leads to universal two-dimensional CFT answers, with the left- and right-moving temperatures appearing appropriately in the entanglement entropy. Altogether, the central charge and left- and right-moving temperatures identified in this way satisfy the Cardy formula and thus reproduce the black hole entropy in the bulk. The central charge we have identified from the entanglement entropy calculation has been previously produced in the literature \cite{Anninos:2008qb} by \emph{demanding} consistency with the Cardy formula. Our approach implements the covariant holographic entanglement entropy proposal and consistency with the Cardy formula is instead a promising output. Taking our results at face value, they seem to suggest that warped CFTs behave like ordinary CFTs in the IR; this matches the intuition garnered from asymptotically warped AdS$_3$ spacetimes in holography, since their deep interiors are AdS$_3$-like for small warping. Our perturbative expansion also shows that there exists nontrivial fiber-coordinate dependence at subleading order in the separation of the fiber coordinate, suggesting that the full theory is not a standard conformal field theory. How to implement a proposal for holographically computing entanglement entropy in asymptotically warped AdS$_3$ spacetimes, without taking an IR limit, remains an open question.

The most immediate way one can make progress on the questions discussed in this paper is by studying the constraints of warped CFT on field-theoretic calculations of entanglement entropy. It has been shown in \cite{Detournay:2012pc} that warped conformal invariance is strongly constraining and allows one to reproduce a Cardy-like formula for the asymptotic growth in the density of states by using the modular covariance of the partition function. As shown in  \cite{Holzhey:1994we}, the calculation of entanglement entropy in the vacuum and finite-temperature states of two-dimensional CFT can be conformally mapped to the calculation of a partition function. The constrained form of the partition function then allows one to write down the universal formulas for two-dimensional CFT. Such a procedure may prove fruitful in the case of warped CFTs as well, although one of the primarily difficulties is due to warped CFTs not having natural Euclidean descriptions. Obtaining a universal entanglement entropy formula for simple states of warped CFTs will allow one to determine if our holographic results are indicating the existence of a second hidden Virasoro algebra or if the infinite-dimensional $U(1)$ is sufficient to constrain the answers in the way we have presented.

It is also interesting to see how far the analogy with two-dimensional CFT can be taken. For example, it is possible that for large separation in the $U(1)$ coordinate, with an appropriate IR limit, the field-theoretic calculation of entanglement entropy in a warped CFT reproduces the CFT result. A simpler question is the constraint on correlation functions: it can be shown \cite{Hofman:2011zj} that left-translation invariance, left-scale invariance, and right-translation invariance constrain the vacuum two-point function of local operators $\phi_i$ to be of the form
\be
\langle \phi_i(x^-,x^+)\phi_j(y^-,y^+)\rangle=\f{f_{ij}(x^--y^-)}{(x^+-y^+)^{\lambda_i+\lambda_j}}\;,
\ee
where $\lambda_i$ is the weight of the operator $\phi_i$. Furthermore, the symmetries are automatically enhanced to an infinite-dimensional left-moving $U(1)$ Kac-Moody algebra and a left-moving Virasoro algebra. If the analogy to two-dimensional CFT is to be taken seriously, these symmetries should provide a constraint on $f_{ij}$ such that in the limit of large separation $x^--y^-$ and in an appropriate infrared regime the answer reduces to that of two-dimensional CFT. Even if this simplification occurs, however, it does not imply that the theory can be described by an ordinary two-dimensional CFT in this regime. The entanglement entropy in the states we have considered and the vacuum two-point function give limited information about the theory and do not elucidate its full dynamics.

Another home for the study of warped AdS$_3$ and warped BTZ black holes is topologically massive gravity, a higher curvature theory of gravity. There exists a proposal for extending the holographic entanglement entropy proposal to this theory \cite{Sun:2008uf}, although there has not been much work in this direction. Given our study of finite-temperature solutions, it is plausible that a proposal for topologically massive gravity which reproduces the CFT$_2$ answer for empty, warped AdS$_3$ will also reproduce the correct answer for the warped BTZ black hole, as shown in Section \ref{finitetemp}.

We have seen that the method of holographically computing entanglement entropy, devised in AdS/CFT, can be adapted to the case of warped AdS$_3$ holography. It provides further evidence that a sharp holographic correspondence can be developed in this context. The perturbative approach we implemented may be a promising way to study entanglement entropy in more general spacetimes continuously connected to AdS$_{d+2}$. It can also be adapted to the NHEK geometry, where one would like to independently deduce $c_L=12J$.

\section*{Acknowledgements}

We would like to acknowledge useful conversations with Tatsuo Azeyanagi, Xi Dong, Michael Gutperle, Sean Hartnoll, Eliot Hijano, Diego Hofman, and Gim Seng Ng. This work has been partially funded by DOE grant DE-FG02-91ER40654. E.S. is supported in part by NSF Grant PHY-0756174 and would like to thank KU Leuven and KITP for their hospitality while part of this work was performed. J.S. is supported in part by NSF Grant PHY-07-57702. D.A. would like to acknowledge the hospitality of the Aspen Center for Physics where part of this work was performed.

\appendix
\section{Geodesics}

\subsection{AdS$_3$ in Poincar\'e fibered coordinates}\label{poinadssols}

There are four solution branches for the geodesics in the background (\ref{fads3}).  To obtain these solutions, one solves equations \eqref{pcphi} and \eqref{pcpsi} for $\dot \psi$ and $\dot \phi$ in terms of $x$, plugs the result back into \eqref{pcE}, and then integrates the resulting equation to obtain
\begin{align}
    x_{c,\pm}(\lambda)
    &=\frac{\ell  c_v}{4 c_{\psi } c_{\phi }\pm 2 \sqrt{c_{\psi }^2 \left(-\left(c_v-4 c_{\phi }^2\right)\right)} \cosh \left(\frac{2 \sqrt{c_v} (\lambda -\text{$\lambda $0})}{\ell }\right)}\;, \label{pxcosh}\\
    x_{s,\pm}(\lambda)
    &=\frac{\ell  c_v}{4 c_{\psi } c_{\phi }\pm 2 \sqrt{c_{\psi }^2 \left(c_v-4 c_{\phi }^2\right)} \sinh \left(\frac{2 \sqrt{c_v} (\lambda -\text{$\lambda $0})}{\ell }\right)}\;. \label{pxsinh}
\end{align}
We typically refer to the solutions \eqref{pxcosh} as the ``cosh-like" branch and to those in \eqref{pxsinh} as the ``sinh-like" branch.  The differences between these branches will be clarified in the next section when we consider global coordinates.  Note that in the process of obtaining these four solutions, one made the assumption that $c_v-4 c_\phi^2<0$ to get the ``cosh-like" branch, while one assumed that $c_v-4c_\phi^2>0$ to obtain the ``sinh-like" branch.  Each of these solutions for $x$ can then be combined with the other conservation equations to find the following corresponding solutions for $\phi$:
\begin{align}
    \phi_{c,\pm}(\lambda)
    &= \widehat\phi_{c,\pm} +2 \ell  \coth ^{-1}\left(\frac{\sqrt{c_v} c_{\psi } \coth \left(\frac{\sqrt{c_v} (\lambda -\text{$\lambda $0})}{\ell }\right)}{2 c_{\psi } c_{\phi }\mp\sqrt{c_{\psi }^2 \left(-\left(c_v-4 c_{\phi }^2\right)\right)}}\right),\\
    \phi_{s,\pm}(\lambda)
    &=\widehat\phi_{s,\pm}\mp  2 \ell  \coth ^{-1}\left(\frac{\sqrt{c_v} c_{\psi }}{\mp2 c_{\psi } c_{\phi } \tanh \left(\frac{\sqrt{c_v} (\lambda -\text{$\lambda $0})}{\ell }\right)+\sqrt{c_{\psi }^2 \left(c_v-4 c_{\phi }^2\right)}}\right),
\end{align}
and the following for $\psi$:
\begin{align}
    \psi_{c,\pm}(\lambda)
    &= \widehat\psi_{c,\pm}+\frac{\ell  \sqrt{c_v} \left(4 c_{\phi } \sqrt{4 c_{\phi }^2-c_v} \sinh \left(\frac{2 \sqrt{c_v} (\lambda -\text{$\lambda $0})}{\ell }\right)+\left(c_v-4 c_{\phi }^2\right) \sinh \left(\frac{4 \sqrt{c_v} (\lambda -\text{$\lambda $0})}{\ell }\right)\right)}{2 c_{\psi } \left(\left(c_v-4 c_{\phi }^2\right) \cosh \left(\frac{4 \sqrt{c_v} (\lambda -\text{$\lambda $0})}{\ell }\right)+c_v+4 c_{\phi }^2\right)}\;,\\
    \psi_{s,\pm}(\lambda)
    &= \widehat\psi_{s,\pm}\pm\frac{\ell  \sqrt{c_v} \left(c_v-4 c_{\phi }^2\right) \cosh \left(\frac{2 \sqrt{c_v} (\lambda -\text{$\lambda $0})}{\ell }\right)}{\pm2 c_{\psi } \left(c_v-4 c_{\phi }^2\right) \sinh \left(\frac{2 \sqrt{c_v} (\lambda -\text{$\lambda $0})}{\ell }\right)+4 c_{\phi } \sqrt{c_{\psi }^2 \left(c_v-4 c_{\phi }^2\right)}}\;.
\end{align}
At first glance one might be concerned about the continuity of the ``cosh-like" branch solutions because of the presence of the function $\coth^{-1}$.  However, we see that the argument of the $\coth^{-1}$ is of the form $\alpha\coth(\beta\lambda+\gamma)$ where $\alpha$, $\beta$, and $\gamma$ are real, and this ensures that the overall solution is continuous.  Similar remarks hold for the sinh branch.

\subsection{AdS$_3$ in global fibered coordinates}\label{globaladssols}
The geodesics of (\ref{globalfiberedads}) are obtained by first solving \eqref{req} to obtain the following four solution branches:
\begin{align}
    r_{\pm,\mathrm c}(\lambda)
    &= \frac{c_uc_\tau}{c_v(\ell/2)^2}\pm \frac{\sqrt{(c_u^2-c_v (\ell/2)^2)(c_\tau^2+c_v(\ell/2)^2)}}{c_v(\ell/2)^2}\cosh\left(\frac{\sqrt{c_v}}{(\ell/2)}(\lambda - \lambda_0)\right), \label{rsolcads}\\
    r_{\pm,\mathrm s}(\lambda)
    &= \frac{c_uc_\tau}{c_v(\ell/2)^2}\mp \frac{\sqrt{-(c_u^2-c_v(\ell/2)^2)(c_\tau^2+c_v(\ell/2)^2)}}{c_v(\ell/2)^2}\sinh\left(\frac{\sqrt{c_v}}{(\ell/2)}(\lambda - \lambda_0)\right) .\label{rsolsads}
\end{align}
The different branches can be clarified by considering the parametrization of the boundary in these coordinates found in Appendix A of \cite{Anninos:2009zi}. The two solutions in the ``cosh-like" branch give geodesics that go from $r=+\infty$ or $r=-\infty$ back to $r=+\infty$ or $r=-\infty$, respectively, depending on the sign of the solution chosen. The two branches in the ``sinh-like" solution correspond to geodesics that go from $r=+\infty$ to $r=-\infty$ or vice versa. In this paper we primarily restrict attention to ``cosh-like" branch solutions. If we define
\begin{align}
    f_c(\lambda) &= e^{\frac{\sqrt{c_v} (\lambda -\text{$\lambda $0})}{(\ell/2)}} h_c\;, \qquad h_c = \sqrt{\left(c_u^2-c_v (\ell/2)^2\right)\left(c_{\tau }^2+c_v (\ell/2)^2\right)}\,,\\
    f_s(\lambda) &= e^{\frac{\sqrt{c_v} (\lambda -\text{$\lambda $0})}{(\ell/2)}} hs\;, \qquad h_s=\sqrt{-\left(c_u^2-c_v (\ell/2)^2\right)\left(c_{\tau }^2+c_v (\ell/2)^2\right)}\,,
\end{align}
then the corresponding solutions for $\tau$ can be written as
\begin{align}
    \tau_{\pm,c}(\lambda)
    &= \widehat \tau_{\pm, c}-\cot ^{-1}\left(\frac{f_c(\lambda)^2\pm 2 c_{\tau } c_u f_c(\lambda)+\left(c_{\tau }^2-c_v (\ell/2)^2\right) \left(c_u^2-c_v (\ell/2)^2\right)}{2 \sqrt{c_v} (\ell/2) \left(c_u^2c_\tau-c_v c_\tau (\ell/2)^2 \pm c_u f_c(\lambda)\right)}\right)\,, \label{tgsc}\\
    \tau_{\pm,s}(\lambda)
    &= \widehat \tau_{\pm, s}-\cot ^{-1}\left(\frac{f_s(\lambda)^2\mp 2 c_{\tau } c_u f_s(\lambda)+\left(c_{\tau }^2-c_v (\ell/2)^2\right) \left(c_u^2-c_v (\ell/2)^2\right)}{2 \sqrt{c_v} (\ell/2) \left(c_u^2c_\tau-c_v c_\tau (\ell/2)^2 \mp c_u f_s(\lambda)\right)}\right)\,. \label{tgss}
\end{align}
The seeming discontinuity of the function arccot$(y)$ at $y=0$ is not important since it can be glued onto arccot$(y)+\pi$ there and continue smoothly to negative values of the argument. This can potentially introduce a shift of $\pi$ into $L_\tau$, which is important and needs to be tracked.\footnote{An easy way to not have to deal tracking constant shifts like this one is to do the naive calculation first and obtain a function of the form $\sin((L_\tau+c)/2)$ in the entanglement entropy answer for the ``cosh-like" branch, with $c$ an overall constant that has not been carefully tracked. Requiring the length to vanish when $L_\tau\rightarrow 0$ now fixes $c=0$.} Now if we define
\begin{align}
    g_{c,\pm}(\lambda) &= c_u \cosh \left(\frac{\sqrt{c_v}}{(\ell/2)}\left(\lambda -\lambda _0\right)\right)\pm (\ell/2) \sqrt{c_v} \sinh \left(\frac{\sqrt{c_v}}{(\ell/2)}\left(\lambda -\lambda _0\right)\right)\,,\\
    g_{s,\pm}(\lambda) &= (\ell/2) \sqrt{c_v} \cosh \left(\frac{\sqrt{c_v}}{(\ell/2)}\left(\lambda -\lambda _0\right)\right)\pm c_u \sinh \left(\frac{\sqrt{c_v}}{(\ell/2)}\left(\lambda -\lambda _0\right)\right)\,,
\end{align}
the solutions for $u$ become
\begin{align}
    u_{\pm,c}(\lambda) &= \widehat u_{\pm ,c} +\frac{1}{2} \log \left(\frac{\left(c_u-(\ell/2) \sqrt{c_v}\right) \left(\left((\ell/2)^2 c_v+c_{\tau }^2\right) g_{c,+}(\lambda)\pm c_{\tau } h_c\right)}{\left((\ell/2) \sqrt{c_v}+c_u\right) \left(\left((\ell/2)^2 c_v+c_{\tau }^2\right) g_{c,-}(\lambda)\pm c_{\tau } h_c\right)}\right)\,, \label{ugsc}\\
    u_{\pm,s}(\lambda) &= \widehat u_{\pm, s} + \frac{1}{2} \log \left(\frac{\left(c_u-(\ell/2) \sqrt{c_v}\right) \left(\left((\ell/2)^2 c_v+c_{\tau }^2\right) g_{s,+}(\lambda)\pm c_{\tau } h_s\right)}{\left((\ell/2) \sqrt{c_v}+c_u\right) \left(\left((\ell/2)^2 c_v+c_{\tau }^2\right) g_{s,-}(\lambda)\mp c_{\tau } h_s\right)}\right)\,. \label{ugss}
\end{align}

\subsection{Warped $\mathrm{AdS}_3$ in global fibered coordinates}\label{globalwadssols}
 To simplify our expressions for geodesics in the background  \eqref{globalfibered} with $a\neq 1$, we define
\begin{align}
     c_1 = -\frac{c_u^2 - a^2c_u^2-(\ell/2)^2a^2c_v}{(\ell/2)^4a^2}, \qquad c_2 = \frac{2}{(\ell/2)^4}c_uc_\tau, \qquad c_3 =\frac{c_u^2-a^2c_\tau^2-(\ell/2)^2a^2c_v}{(\ell/2)^4a^2},\label{constants}
\end{align}
This illuminates the general form of \eqref{req}:
\begin{align}
    \dot{r}^2 = c_1r^2 -c_2 r - c_3\, .\label{reqsimp}
\end{align}
The general solution with $c_v>0$ has four branches depending on the sign of the combination $c_2^2+4c_1c_3$.  We ignore the case $c_2^2+4c_1c_3=0$ which yields an exponentially decaying solution.  For $c_2^2+4c_1c_3>0$, there are two ``cosh-like" branches $r_{\pm, c}$ and for $c_2^2+4c_1c_3<0$, there are two ``sinh-like" branches $r_{\pm, s}$:
\begin{align}
    r_{\pm,\mathrm c}(\lambda)
    &= \frac{c_2}{2c_1}\pm \sqrt{\left(\frac{c_2}{2c_1}\right)^2+\frac{c_3}{c_1}}\cosh\big(\sqrt{c_1}(\lambda - \lambda_0)\big) \label{rsolc}\,,\\
    r_{\pm, \mathrm s}(\lambda)
    &= \frac{c_2}{2c_1}\mp \sqrt{-\left(\frac{c_2}{2c_1}\right)^2-\frac{c_3}{c_1}}\sinh\big(\sqrt{c_1}(\lambda - \lambda_0)\big) \label{rsols}\,.
\end{align}
Comparing these solutions to \eqref{rsolcads} and \eqref{rsolsads}, we find the same qualitative behavior in $r(\lambda)$ for both the warped and non-warped cases.  Moreover, setting $a=1$ in these warped solutions yields precisely \eqref{rsolcads} and \eqref{rsolsads}, as one would expect since the form of the $r$ equation is left unaltered by non-trivial warping.  The form of the $\tau$ equation is unaltered by warping, so we expect the corresponding solution to be of the same form.  However, notice that the last term in \eqref{taueq} only appears in the case $a\neq 1$ where there is non-trivial warping, and this changes the qualitative behavior of the solutions.  In particular, manipulating the second and third terms allows one to write the $u$ equation as
\begin{align}
    \dot{u}
    &= \frac{c_u}{(\ell/2)^2a^2}-\frac{c_u}{(\ell/2)^2}\frac{r^2}{r^2+1}+\frac{c_\tau}{(\ell/2)^2}\frac{r}{r^2+1} \label{ueq2},
\end{align}
from which it becomes clear that integration of the first term with respect to $\lambda$ leads to a term in the solution for $u$ that diverges linearly with $\lambda$. This is just like the AdS$_2\times\mathbb{R}$ case.

\subsubsection{$c_u=0$}\label{cueq0}
A simple limit in which we can compute the length of the geodesic in terms of separations in $\tau$ and $u$ at large $r$ is for $c_u=0$. In this case, the equations of motion become $a$-independent and we are forced onto the ``sinh-like"  branch. The answer is then just given by the answer for AdS$_3$ in global fibered coordinates with $c_u=0$:
\begin{align}
\textrm{Length}\sim \log\left(\sqrt{\cos\left(\frac{L_\tau}{2}\right)}\,\frac{\ell}{\epsilon}\right).
\end{align}
It is not a problem that the argument of the log does not begin to vanish in the $L_\tau=\tau(\lambda_\infty)-\tau(-\lambda_\infty)\rightarrow 0$ limit, since we are on the ``sinh-like" branch and so $r(\lambda_\infty)\neq r(-\lambda_\infty)$. Thus, the endpoints remain well-separated in the limit $L_\tau\rightarrow 0$.

\section{$\mathrm{AdS}_3$ entanglement entropy via coordinate transformations}\label{appendix}
For completeness, in this section we show how one can translate from entanglement entropy answers in Poincar\'e coordinates to global coordinates, or vice versa, by performing the appropriate coordinate transformations. The only point one needs to be careful about is the mapping of the UV cutoff. As an illustrastive example, we will begin with showing how to get the answer in global coordinates from the answer in the Poincar\'e patch, within which it is easiest to compute. We then show how to go from Poincar\'e fibered coordinates to global fibered coordinates. Such methods will come in handy when we go from warped AdS$_3$ to the warped BTZ black hole in Section \ref{finitetemp}.

\subsection{Global coordinates from Poincar\'e patch}\label{globalpoincare}

Recall that $\mathrm{AdS}_3$ can be defined as an embedded submanifold of $\reals^{2,2}$ defined by the constraint
\begin{align}
  X_0^2-X_1^2-X_2^2+X_3^2 = 1
\end{align}
where $X_0,X_1,X_2,X_3$ are the standard coordinates on $\reals^{2,2}$.  The embedding coordinates for global $\mathrm{AdS}_3$ are
\begin{align}
  X_0&=\ell \cosh\rho \cos t_g\,,\qquad X_1= \ell\sinh\rho\sin\theta_g\,, \\
  X_2&=\ell\sinh\rho\cos\theta_g\,, \qquad X_3=\ell\cosh\rho\sin t_g\,,
\end{align}
while for Poincar\'e AdS$_3$ they are
\begin{align}
  X_0&=\f{1}{2z}(z^2+\ell^2+x^2-t^2)\,,\qquad X_1=\frac{\ell x}{z}\;, \\ X_2&=\f{1}{2z}(z^2-\ell^2+x^2-t^2)\,,\qquad X_3= \frac{\ell  t}{z}\;.
\end{align}
To get from global coordinates to Poincar\'e coordinates we use the transformations
\begin{align*}
  \f{X_0^2+X_3^2}{\ell^2}&=\cosh^2\rho=\f{1}{4\ell^2z^2}(z^2+\ell^2+x^2-t^2)^2+\f{t^2}{z^2}\;,\\
  \f{X_3}{X_0}&=\tan t_g=\f{2\ell t}{(z^2+\ell^2+x^2-t^2)}\;,\\
  \f{X_1}{X_2}&=\tan\theta_g=\f{2\ell x}{(z^2-\ell^2+x^2-t^2)}\;.
\end{align*}
The inverse transformations are given by
\begin{align*}
\f{\ell^2}{X_0-X_2}&=z=\f{\ell}{\cosh\rho\cos t_g-\sinh\rho\cos\theta_g}\;,\\
\f{\ell X_1}{X_0-X_2}&=x=\f{\ell\sinh\rho\sin\theta}{\cosh\rho\cos t_g-\sinh\rho\cos\theta_g}\;,\\
\f{\ell X_3}{X_0-X_2}&=t=\f{\ell\cosh\rho\sin t_g}{\cosh\rho\cos t_g-\sinh\rho\cos\theta_g}\;.
\end{align*}
Using either set of relationships, we can see that for $t=t_g=0$ and $z=0$, $\rho=\infty$, we get $x=\ell\sin\theta_g/(1-\cos\theta_g)$. We want to show that $L_x/\ep_P=(x_2-x_1)/\ep_P$, when written in terms of $L_{\theta}$, is $L_x/\ep_P\propto \sin( L_{\theta}/2)/\ep_g$. Our answer for the length of the curve in Poincar\'e coordinates is
\begin{align}
\f{c}{6}\log\left(\f{L_x}{z_1}\right)+\f{c}{6}\log\left(\f{L_x}{z_2}\right)=\f{c}{3}\,\log\left(\f{L_x}{\sqrt{z_1 z_2}}\right)
\end{align}
where we have picked two different endpoints $z_1$ and $z_2$ for the curve. Using $z=\f{2\ell e^{-\rho}}{1-\cos\theta_g}$ as the asymptotic coordinate transformation between the coordinates gives
\begin{align}
\f{L_x}{\sqrt{z_1z_2}}&=\left(\f{\sin{\theta_{g,2}}}{1-\cos\theta_{g,2}}-\f{\sin{\theta_{g,1}}}{1-\cos\theta_{g,1}}\right)
\f{\sqrt{(1-\cos\theta_{g,1})(1-\cos\theta_{g,2})}}{2e^{-\rho}}\\
&=\f{\sin\left(\f{L_{\theta}}{2}\right)}{e^{-\rho}}\;,
\end{align}
where we have picked $\rho_1=\rho_2=\rho$ to fix to a constant cutoff surface. We then use $e^{-\rho}\sim\f{a}{L}\sim\ep$ and recover
\begin{align}
S_{global}=\f{c}{3}\,\log\f{\sin\f{L_{\theta}}{2}}{\ep}=\f{c}{3}\,\log\f{\sin (l\pi/L)}{\ep}
\end{align}
upon identifying $L_{\theta}=\theta_{g,2}-\theta_{g,1}=2\pi l/L$ for total circumference $L$.

\subsection{Global fibered from Poincar\'e fibered}\label{fiberedcoords}
We now map the Poincar\'e fibered answer onto the global fibered answer. The coordinate transformations between these two metrics are
\begin{align*}
  \phi&=\ell\log\left(\f{e^{\sigma}\cot(\tau/2)-1}{e^{\sigma}\cot(\tau/2)+1}e^u\right),\quad
  \psi=\f{\cosh\sigma\sin\tau}{\sinh\sigma+\cosh\sigma\cos\tau}\;,\quad
  x=\f{1}{\sinh\sigma+\cosh\sigma\cos\tau}\;,
\end{align*}
which asymptotically ($\sigma\rightarrow\infty$) become
\begin{align*}
  \phi&=\ell u\,,\qquad
  \psi=\tan(\tau/2)\, ,\qquad
  x=\f{2e^{-\sigma}}{1+\cos\tau}\, .
\end{align*}
From these relations we see that
\begin{align*}
L_{\phi}&=\ell L_u\, ,\qquad
L_{\psi}=\tan(\tau_2/2)-\tan(\tau_1/2)\, ,
\end{align*}
and
\begin{align*}
  \f{1}{\sqrt{x_1x_2}}&=\f{\sqrt{(1+\cos\tau_1)(1+\cos\tau_2)}}{2e^{-\sigma}}\, ,
\end{align*}
since we will be assuming we are at different points $x_1$, $x_2$ at the two ends of the curve in Poincar\'e fibered coordinates, whereas for global fibered coordinates we will assume we are at the same $\sigma$ coordinate at both endpoints of the curve. Our Poincar\'e fibered answer was found to be
\begin{align}
S=\f{c}{3}\,\log\left(\f{\sqrt{L_{\psi}\ell\sinh\f{L_{\phi}}{2\ell}}}{\ep}\right),
\end{align}
where to get here the cutoff relation $x\sim\ep^2/\ell$ was employed. Translating back, we find
\begin{align*}
\ep\rightarrow\sqrt{\ep_1\ep_2}=\sqrt{\ell\sqrt{x_1x_2}}
\end{align*}
gives us
\begin{align*}
S&=\f{c}{3}\,\log\sqrt{\f{(\tan(\tau_2/2)-\tan(\tau_1/2))\sqrt{(1+\cos\tau_1)(1+\cos\tau_2)}}{2e^{-2\sigma}}\sinh\f{L_{u}}{2}}\\
&=\f{c}{3}\,\log\sqrt{\f{\sin(L_{\tau}/2)\sinh(L_{u}/2)}{e^{-2\sigma}}}=\f{c}{3}\,\log\left(\f{1}{\epsilon}\sqrt{\sin\left(\f{L_{\tau}}{2}\right)\sinh\left(\f{L_{u}}{2}\right)}\right)
\end{align*}
where we have used the relation $\ep\sim e^{-\sigma}$ for what is now a dimensionless UV cutoff. This is precisely the answer for global fibered coordinates \eqref{globalfiber}.
\\


\begin{thebibliography}{99}

\bibitem{ds}
  Holographic aspects of FRW/de Sitter geometries include:

A.~Strominger,
  ``The dS / CFT correspondence,''
  JHEP {\bf 0110}, 034 (2001)
  [hep-th/0106113].

 D.~Anninos, T.~Hartman and A.~Strominger,
  ``Higher Spin Realization of the dS/CFT Correspondence,''
  arXiv:1108.5735 [hep-th].

  M.~Alishahiha, A.~Karch, E.~Silverstein and D.~Tong,
  ``The dS/dS correspondence,''
  AIP Conf.\ Proc.\  {\bf 743}, 393 (2005)
  [hep-th/0407125].

  X.~Dong, B.~Horn, E.~Silverstein and G.~Torroba,
  ``Micromanaging de Sitter holography,''
  Class.\ Quant.\ Grav.\  {\bf 27}, 245020 (2010)
  [arXiv:1005.5403 [hep-th]].

    D.~Anninos, S.~A.~Hartnoll and D.~M.~Hofman,
  ``Static Patch Solipsism: Conformal Symmetry of the de Sitter Worldline,''
  Class.\ Quant.\ Grav.\  {\bf 29}, 075002 (2012)
  [arXiv:1109.4942 [hep-th]].

    B.~Freivogel, Y.~Sekino, L.~Susskind and C.~-P.~Yeh,
  ``A Holographic framework for eternal inflation,''
  Phys.\ Rev.\ D {\bf 74}, 086003 (2006)
  [hep-th/0606204].

  P.~McFadden and K.~Skenderis,
  ``Holography for Cosmology,''
  Phys.\ Rev.\ D {\bf 81}, 021301 (2010)
  [arXiv:0907.5542 [hep-th]].

  E.~Shaghoulian,
  ``FRW cosmologies and hyperscaling-violating geometries: higher curvature corrections, ultrametricity, Q-space/QFT duality, and a little string theory,''
  arXiv:1308.1095 [hep-th].

  D.~Anninos,
  ``De Sitter Musings,''
  Int.\ J.\ Mod.\ Phys.\ A {\bf 27}, 1230013 (2012)
  [arXiv:1205.3855 [hep-th]].




\bibitem{nhek}
  Holographic aspects of NHEK/warped AdS$_3$ geometries include:

  J.~M.~Bardeen and G.~T.~Horowitz,
  ``The Extreme Kerr throat geometry: A Vacuum analog of AdS(2) x S**2,''
  Phys.\ Rev.\ D {\bf 60}, 104030 (1999)
  [hep-th/9905099].

    D.~Anninos, W.~Li, M.~Padi, W.~Song and A.~Strominger,
  ``Warped AdS(3) Black Holes,''
  JHEP {\bf 0903}, 130 (2009)
  [arXiv:0807.3040 [hep-th]].

  M.~Guica, T.~Hartman, W.~Song and A.~Strominger,
  ``The Kerr/CFT Correspondence,''
  Phys.\ Rev.\ D {\bf 80}, 124008 (2009)
  [arXiv:0809.4266 [hep-th]].




\bibitem{schro}
  Holographic aspects of Schrodinger and Lifshitz geometries include:

  K.~Balasubramanian and J.~McGreevy,
  ``Gravity duals for non-relativistic CFTs,''
  Phys.\ Rev.\ Lett.\  {\bf 101}, 061601 (2008)
  [arXiv:0804.4053 [hep-th]].

  D.~T.~Son,
  ``Toward an AdS/cold atoms correspondence: A Geometric realization of the Schrodinger symmetry,''
  Phys.\ Rev.\ D {\bf 78}, 046003 (2008)
  [arXiv:0804.3972 [hep-th]].

  S.~Kachru, X.~Liu and M.~Mulligan,
  ``Gravity Duals of Lifshitz-like Fixed Points,''
  Phys.\ Rev.\ D {\bf 78}, 106005 (2008)
  [arXiv:0808.1725 [hep-th]].




\bibitem{Nutku:1993eb}
  Y.~Nutku,
  ``Exact solutions of topologically massive gravity with a cosmological constant,''
  Class.\ Quant.\ Grav.\  {\bf 10}, 2657 (1993).

\bibitem{gurses}
M. G\"{u}rses, ``Perfect fluid sources in 2+1 dimensions,"
Class. Quant. Grav. 11, 2585 (1994).

\bibitem{Moussa:2003fc}
  K.~A.~Moussa, G.~Clement and C.~Leygnac,
  ``The Black holes of topologically massive gravity,''
  Class.\ Quant.\ Grav.\  {\bf 20}, L277 (2003)
  [gr-qc/0303042].

\bibitem{Bouchareb:2007yx}
  A.~Bouchareb and G.~Clement,
  ``Black hole mass and angular momentum in topologically massive gravity,''
  Class.\ Quant.\ Grav.\  {\bf 24}, 5581 (2007)
  [arXiv:0706.0263 [gr-qc]].

\bibitem{Anninos:2008qb}
  D.~Anninos,
  ``Hopfing and Puffing Warped Anti-de Sitter Space,''
  JHEP {\bf 0909}, 075 (2009)
  [arXiv:0809.2433 [hep-th]].

\bibitem{Orlando:2010ay}
  D.~Orlando and L.~I.~Uruchurtu,
  ``Warped anti-de Sitter spaces from brane intersections in type II string theory,''
  JHEP {\bf 1006}, 049 (2010)
  [arXiv:1003.0712 [hep-th]].

\bibitem{Song:2011sr}
  W.~Song and A.~Strominger,
  ``Warped AdS3/Dipole-CFT Duality,''
  JHEP {\bf 1205}, 120 (2012)
  [arXiv:1109.0544 [hep-th]].

\bibitem{Detournay:2010rh}
  S.~Detournay, D.~Israel, J.~M.~Lapan and M.~Romo,
  ``String Theory on Warped $AdS_{3}$ and Virasoro Resonances,''
  JHEP {\bf 1101}, 030 (2011)
  [arXiv:1007.2781 [hep-th]].

\bibitem{ElShowk:2011cm}
  S.~El-Showk and M.~Guica,
  ``Kerr/CFT, dipole theories and nonrelativistic CFTs,''
  JHEP {\bf 1212}, 009 (2012)
  [arXiv:1108.6091 [hep-th]].

\bibitem{Azeyanagi:2012zd}
  T.~Azeyanagi, D.~M.~Hofman, W.~Song and A.~Strominger,
  ``The Spectrum of Strings on Warped AdS$_3$ x $S^3$,''
  JHEP {\bf 1304}, 078 (2013)
  [arXiv:1207.5050 [hep-th]].

\bibitem{Karndumri:2013dca} 
  P.~Karndumri and E.~O Colgain,
  ``3D Supergravity from wrapped D3-branes,''
  JHEP {\bf 1310}, 094 (2013)
  [arXiv:1307.2086 [hep-th]].

\bibitem{Hofman:2011zj}
  D.~M.~Hofman and A.~Strominger,
  ``Chiral Scale and Conformal Invariance in 2D Quantum Field Theory,''
  Phys.\ Rev.\ Lett.\  {\bf 107}, 161601 (2011)
  [arXiv:1107.2917 [hep-th]].

\bibitem{Detournay:2012pc}
  S.~Detournay, T.~Hartman and D.~M.~Hofman,
  ``Warped Conformal Field Theory,''
  Phys.\ Rev.\ D {\bf 86}, 124018 (2012)
  [arXiv:1210.0539 [hep-th]].

\bibitem{Chen:2009rf}
  B.~Chen and Z.~-b.~Xu,
  ``Quasinormal modes of warped AdS(3) black holes and AdS/CFT correspondence,''
  Phys.\ Lett.\ B {\bf 675}, 246 (2009)
  [arXiv:0901.3588 [hep-th]].

\bibitem{Chen:2009hg}
  B.~Chen and Z.~-b.~Xu,
  ``Quasi-normal modes of warped black holes and warped AdS/CFT correspondence,''
  JHEP {\bf 0911}, 091 (2009)
  [arXiv:0908.0057 [hep-th]].

\bibitem{Chen:2009cg}
  B.~Chen, B.~Ning and Z.~-b.~Xu,
  ``Real-time correlators in warped AdS/CFT correspondence,''
  JHEP {\bf 1002}, 031 (2010)
  [arXiv:0911.0167 [hep-th]].

\bibitem{Anninos:2009jt}
  D.~Anninos,
  ``Sailing from Warped AdS(3) to Warped dS(3) in Topologically Massive Gravity,''
  JHEP {\bf 1002}, 046 (2010)
  [arXiv:0906.1819 [hep-th]].

\bibitem{Anninos:2010gh}
   D.~Anninos and T.~Anous,
  ``A de Sitter Hoedown,''
  JHEP {\bf 1008}, 131 (2010)
  [arXiv:1002.1717 [hep-th]].

\bibitem{Compere:2008cv}
  G.~Compere and S.~Detournay,
  ``Semi-classical central charge in topologically massive gravity,''
  Class.\ Quant.\ Grav.\  {\bf 26}, 012001 (2009)
  [Erratum-ibid.\  {\bf 26}, 139801 (2009)]
  [arXiv:0808.1911 [hep-th]].

\bibitem{Compere:2009zj}
  G.~Compere and S.~Detournay,
  ``Boundary conditions for spacelike and timelike warped AdS$_3$ spaces in topologically massive gravity,''
  JHEP {\bf 0908}, 092 (2009)
  [arXiv:0906.1243 [hep-th]].

\bibitem{Hubeny:2007xt}
  V.~E.~Hubeny, M.~Rangamani and T.~Takayanagi,
  ``A Covariant holographic entanglement entropy proposal,''
  JHEP {\bf 0707}, 062 (2007)
  [arXiv:0705.0016 [hep-th]].

\bibitem{Ryu:2006bv}
  S.~Ryu and T.~Takayanagi,
  ``Holographic derivation of entanglement entropy from AdS/CFT,''
  Phys.\ Rev.\ Lett.\  {\bf 96}, 181602 (2006)
  [hep-th/0603001].

\bibitem{Ryu:2006ef}
  S.~Ryu and T.~Takayanagi,
  ``Aspects of Holographic Entanglement Entropy,''
  JHEP {\bf 0608}, 045 (2006)
  [hep-th/0605073].


\bibitem{Bengtsson:2005zj}
  I.~Bengtsson and P.~Sandin,
  Class.\ Quant.\ Grav.\  {\bf 23}, 971 (2006)
  [gr-qc/0509076].

\bibitem{Horava:2009vy}
  P.~Horava and C.~M.~Melby-Thompson,
  ``Anisotropic Conformal Infinity,''
  Gen.\ Rel.\ Grav.\  {\bf 43}, 1391 (2011)
  [arXiv:0909.3841 [hep-th]].

\bibitem{Anninos:2009zi}
  D.~Anninos, M.~Esole and M.~Guica,
  ``Stability of warped AdS(3) vacua of topologically massive gravity,''
  JHEP {\bf 0910}, 083 (2009)
  [arXiv:0905.2612 [hep-th]].

\bibitem{Compere:2013bya}
  G.~Compere, W.~Song and A.~Strominger,
  ``New Boundary Conditions for AdS3,''
  arXiv:1303.2662 [hep-th].

\bibitem{Duff:1998cr}
  M.~J.~Duff, H.~Lu and C.~N.~Pope,
  ``AdS(3) x S**3 (un)twisted and squashed, and an O(2,2,Z) multiplet of dyonic strings,''
  Nucl.\ Phys.\ B {\bf 544}, 145 (1999)
  [hep-th/9807173].

\bibitem{Israel:2004vv}
  D.~Israel, C.~Kounnas, D.~Orlando and P.~M.~Petropoulos,
  ``Electric/magnetic deformations of S**3 and AdS(3), and geometric cosets,''
  Fortsch.\ Phys.\  {\bf 53}, 73 (2005)
  [hep-th/0405213].

\bibitem{Headrick:2010zt}
  M.~Headrick,
  ``Entanglement Renyi entropies in holographic theories,''
  Phys.\ Rev.\ D {\bf 82}, 126010 (2010)
  [arXiv:1006.0047 [hep-th]].

\bibitem{Hartman:2013mia}
  T.~Hartman,
  ``Entanglement Entropy at Large Central Charge,''
  arXiv:1303.6955 [hep-th].

\bibitem{Faulkner:2013yia}
  T.~Faulkner,
  ``The Entanglement Renyi Entropies of Disjoint Intervals in AdS/CFT,''
  arXiv:1303.7221 [hep-th].

\bibitem{Casini:2011kv}
  H.~Casini, M.~Huerta and R.~C.~Myers,
  ``Towards a derivation of holographic entanglement entropy,''
  JHEP {\bf 1105}, 036 (2011)
  [arXiv:1102.0440 [hep-th]].

\bibitem{Lewkowycz:2013nqa}
  A.~Lewkowycz and J.~Maldacena,
  ``Generalized gravitational entropy,''
  arXiv:1304.4926 [hep-th].

\bibitem{Barrella:2013wja}
  T.~Barrella, X.~Dong, S.~A.~Hartnoll and V.~L.~Martin,
  ``Holographic entanglement beyond classical gravity,''
  arXiv:1306.4682 [hep-th].

\bibitem{Faulkner:2013ana}
  T.~Faulkner, A.~Lewkowycz and J.~Maldacena,
  ``Quantum corrections to holographic entanglement entropy,''
  arXiv:1307.2892 [hep-th].

\bibitem{Hung:2011xb}
  L.~-Y.~Hung, R.~C.~Myers and M.~Smolkin,
  ``On Holographic Entanglement Entropy and Higher Curvature Gravity,''
  JHEP {\bf 1104}, 025 (2011)
  [arXiv:1101.5813 [hep-th]].

\bibitem{Bhattacharyya:2013jma} 
  A.~Bhattacharyya, A.~Kaviraj and A.~Sinha,
  ``Entanglement entropy in higher derivative holography,''
  JHEP {\bf 1308}, 012 (2013)
  [arXiv:1305.6694 [hep-th]].

\bibitem{Fursaev:2013fta} 
  D.~V.~Fursaev, A.~Patrushev and S.~N.~Solodukhin,
  ``Distributional Geometry of Squashed Cones,''
  arXiv:1306.4000 [hep-th].

\bibitem{Bhattacharyya:2013gra} 
  A.~Bhattacharyya, M.~Sharma and A.~Sinha,
  ``On generalized gravitational entropy, squashed cones and holography,''
  arXiv:1308.5748 [hep-th].




\bibitem{Sun:2008uf}
  J.~-R.~Sun,
  ``Note on Chern-Simons Term Correction to Holographic Entanglement Entropy,''
  JHEP {\bf 0905}, 061 (2009)
  [arXiv:0810.0967 [hep-th]].

\bibitem{Ammon:2013hba}
  M.~Ammon, A.~Castro and N.~Iqbal,
  ``Wilson Lines and Entanglement Entropy in Higher Spin Gravity,''
  arXiv:1306.4338 [hep-th].

\bibitem{deBoer:2013vca}
  J.~de Boer and J.~I.~Jottar,
  ``Entanglement Entropy and Higher Spin Holography in AdS$_3$,''
  arXiv:1306.4347 [hep-th].

\bibitem{Callan:2012ip}
  R.~Callan, J.~-Y.~He and M.~Headrick,
  ``Strong subadditivity and the covariant holographic entanglement entropy formula,''
  JHEP {\bf 1206}, 081 (2012)
  [arXiv:1204.2309 [hep-th]].

\bibitem{Wall:2012uf}
  A.~C.~Wall,
  ``Maximin Surfaces, and the Strong Subadditivity of the Covariant Holographic Entanglement Entropy,''
  arXiv:1211.3494 [hep-th].

\bibitem{Hubeny:2013gta}
  V.~E.~Hubeny, H.~Maxfield, M.~Rangamani and E.~Tonni,
  ``Holographic entanglement plateaux,''
  arXiv:1306.4004 [hep-th].

\bibitem{Ogawa:2011bz}
  N.~Ogawa, T.~Takayanagi and T.~Ugajin,
  ``Holographic Fermi Surfaces and Entanglement Entropy,''
  JHEP {\bf 1201}, 125 (2012)
  [arXiv:1111.1023 [hep-th]].

\bibitem{Shaghoulian:2011aa}
  E.~Shaghoulian,
  ``Holographic Entanglement Entropy and Fermi Surfaces,''
  JHEP {\bf 1205}, 065 (2012)
  [arXiv:1112.2702 [hep-th]].

\bibitem{vanRees:2011fr}
  B.~C.~van Rees,
  ``Holographic renormalization for irrelevant operators and multi-trace counterterms,''
  JHEP {\bf 1108}, 093 (2011)
  [arXiv:1102.2239 [hep-th]].

\bibitem{vanRees:2011ir}
  B.~C.~van Rees,
  ``Irrelevant deformations and the holographic Callan-Symanzik equation,''
  JHEP {\bf 1110}, 067 (2011)
  [arXiv:1105.5396 [hep-th]].

\bibitem{Korovin:2013bua}
  Y.~Korovin, K.~Skenderis and M.~Taylor,
  ``Lifshitz as a deformation of Anti-de Sitter,''
  JHEP {\bf 1308}, 026 (2013)
  [arXiv:1304.7776 [hep-th]].

\bibitem{Susskind:1998dq}
  L.~Susskind and E.~Witten,
  ``The Holographic bound in anti-de Sitter space,''
  hep-th/9805114.


\bibitem{fgexpansion}
C.~Fefferman and C.~R.~Graham,
``Conformal Invariants," Elie Cartan et les Mathematiques d'Aujourdui, Asterisque 95 (1985)

\bibitem{Balasubramanian:1999re}
  V.~Balasubramanian and P.~Kraus,
  ``A Stress tensor for Anti-de Sitter gravity,''
  Commun.\ Math.\ Phys.\  {\bf 208}, 413 (1999)
  [hep-th/9902121].

\bibitem{de Haro:2000xn}
  S.~de Haro, S.~N.~Solodukhin and K.~Skenderis,
  ``Holographic reconstruction of space-time and renormalization in the AdS / CFT correspondence,''
  Commun.\ Math.\ Phys.\  {\bf 217}, 595 (2001)
  [hep-th/0002230].

\bibitem{Balasubramanian:2009bg}
  V.~Balasubramanian, J.~de Boer, M.~M.~Sheikh-Jabbari and J.~Simon,
  ``What is a chiral 2d CFT? And what does it have to do with extremal black holes?,''
  JHEP {\bf 1002}, 017 (2010)
  [arXiv:0906.3272 [hep-th]].

\bibitem{Coussaert:1994tu}
  O.~Coussaert and M.~Henneaux,
  ``Selfdual solutions of (2+1) Einstein gravity with a negative cosmological constant,''
  In *Teitelboim, C. (ed.): The black hole* 25-39
  [hep-th/9407181].

\bibitem{Balasubramanian:2003kq}
  V.~Balasubramanian, A.~Naqvi and J.~Simon,
  ``A Multiboundary AdS orbifold and DLCQ holography: A Universal holographic description of extremal black hole horizons,''
  JHEP {\bf 0408}, 023 (2004)
  [hep-th/0311237].

\bibitem{Azeyanagi:2007bj}
  T.~Azeyanagi, T.~Nishioka and T.~Takayanagi,
  ``Near Extremal Black Hole Entropy as Entanglement Entropy via AdS(2)/CFT(1),''
  Phys.\ Rev.\ D {\bf 77}, 064005 (2008)
  [arXiv:0710.2956 [hep-th]].

\bibitem{Morrison:2012iz}
  I.~A.~Morrison and M.~M.~Roberts,
  ``Mutual information between thermo-field doubles and disconnected holographic boundaries,''
  arXiv:1211.2887 [hep-th].

\bibitem{Compere:2007in}
  G.~Compere and S.~Detournay,
  ``Centrally extended symmetry algebra of asymptotically Godel spacetimes,''
  JHEP {\bf 0703}, 098 (2007)
  [hep-th/0701039].

\bibitem{Anninos:2008fx}
    D.~Anninos, W.~Li, M.~Padi, W.~Song and A.~Strominger,
  ``Warped AdS(3) Black Holes,''
  JHEP {\bf 0903}, 130 (2009)
  [arXiv:0807.3040 [hep-th]].

\bibitem{Levi:2009az}
  T.~S.~Levi, J.~Raeymaekers, D.~Van den Bleeken, W.~Van Herck and B.~Vercnocke,
  ``Godel space from wrapped M2-branes,''
  JHEP {\bf 1001}, 082 (2010)
  [arXiv:0909.4081 [hep-th]].

\bibitem{Banados:2005da}
  M.~Banados, G.~Barnich, G.~Compere and A.~Gomberoff,
  ``Three dimensional origin of Godel spacetimes and black holes,''
  Phys.\ Rev.\ D {\bf 73}, 044006 (2006)
  [hep-th/0512105].

\bibitem{Holzhey:1994we}
  C.~Holzhey, F.~Larsen and F.~Wilczek,
  ``Geometric and renormalized entropy in conformal field theory,''
  Nucl.\ Phys.\ B {\bf 424}, 443 (1994)
  [hep-th/9403108].

\end{thebibliography}
\end{document}